\documentclass{aa}
\usepackage{txfonts}
\usepackage{psfig}
\usepackage{graphicx}
\usepackage{natbib}

\begin{document}

\def\eps{\varepsilon}
\def\aap{A\&A}
\def\apj{ApJ}
\def\apjs{ApJS}
\def\apjl{ApJL}
\def\apss{Ap\&SS}
\def\mnras{MNRAS}
\def\aj{AJ}
\def\nat{Nature}
\def\aaps{A\&A Supp.}
\def\prd{Phys. Rev. D}

\def\prl{Phys. Rev. Lett.}
\def\araa{ARA\&A}       


\title{The imprint of cosmological hydrogen recombination lines  
on the power spectrum of the CMB}
\titlerunning{Cosmological hydrogen recombination lines and CMB}
\subtitle{}

\author{J.~A. Rubi\~no-Mart\'{\i}n\inst{1,2}
\and   C. Hern\'andez-Monteagudo\inst{1}\thanks{\emph{Present
address:} Dept. of Physics \& Astronomy, Univ. of Pennsylvania, 
209 South 33rd Str., Philadelphia, PA 19104-6396, USA.}
\and   R.~A. Sunyaev\inst{1,3}
       }
\authorrunning{J. A. Rubi\~no-Mart\'{\i}n}
\offprints{J.A.Rubi\~no-Mart\'{\i}n, \email{jalberto@iac.es}}

\institute {
Max-Planck-Institut f\"{u}r Astrophysik,
Karl-Schwarzschild-Str.1, Postfach 1317, 85741 Garching,
Germany
\and 
Instituto de Astrof\'{\i}sica de Canarias, 38200 La Laguna,
  Tenerife, Spain. 
\and
Space Research Institute (IKI), Russian Academy of Sciences,
Moscow, Russia
}

\date{Received date..............; accepted date................}

\abstract{
We explore the imprint of the cosmological hydrogen recombination
lines on the power spectrum of the cosmic microwave background (CMB). 
In particular, we focus on the three strongest lines for
the Balmer, Paschen and Brackett series of hydrogen. 
We expect changes in the angular
power spectrum due to these lines of about $0.3 \mu K$ 
for the H$\alpha$ line, being maximum at small angular scales 
($\ell \approx 870$).
The morphology of the signal is very rich. It leads to relatively
narrow spectral features ($\Delta \nu / \nu \sim 10^{-1}$), 
with several regions in the power spectrum showing a characteristic change
of sign of the effect as we probe different redshifts or different
multipoles by measuring the power spectrum at different frequencies.
It also has a very peculiar 
dependence on the multipole scale, connected with the details of the
transfer function at the epoch of scattering (see movies illustrating
the effect at {\rm http://www.mpa$-$garching.mpg.de/$\sim$jalberto}). 
In order to compute the optical depths for these transitions, 
we have numerically evolved the populations of the levels of the
hydrogen atom during recombination, simultaneously treating the
evolution of helium. For the hydrogen atom, we follow 
each angular momentum state separately, up to the level $n=10$.
Foregrounds and other frequency dependent contaminants such as Rayleigh
scattering may be a important limitation for these measurements,
although the peculiar frequency and angular dependences of the effect
that we are discussing might permit us to separate it. 
Detection of this signal using future narrow-band spectral observations can
give information about the details of how the cosmic recombination
proceeds, and how Silk damping operates during recombination. 

}
\maketitle

\section{Introduction}

Existing and planned experiments devoted to the
measurements of Cosmic Microwave Background (CMB) 
angular fluctuations will reach an unprecedent high
sensitivity of measurements. 
Planck, SPT, ACT and QUEST  will 
achieve sensitivities several orders
of magnitude higher than the amplitude of 
the acoustic peaks which were the dream of
theorists in the seventies 
\citep{SZ70,1970ApJ...162..815P,1978SvA....22..523D}, 
and are observed with good precision now by 
Boomerang, Maxima, DASI, VSA, CBI, and more recently with very
high accuracy by WMAP \citep{2003ApJS..148....1B}.

The experimental progress demonstrates
that we are entering the era of "precision
cosmology", and many of the effects that were
obvious for theoreticians but of small 
amplitude become now accessible to experimentalists.

One of such problems is the direct detection 
of the hydrogen or helium lines from the
epoch of recombination of hydrogen in the
Universe at redshift $z \sim 1000$.
In Russia, all the study of cosmological recombination grew up
from the question of Vladimir Kurt: ``where are the Ly-$\alpha$ line photons 
from recombination in the Universe? ''. This question 
forced 
\citet[hereafter, ZKS68]{1968ZhETF..55..278Z} to study
the detailed physics of recombination, and to show
that two-photon decay of 2s level of hydrogen is more 
important than the escape of redshifted Ly-$\alpha$ photons due
to expansion of the Universe 
in determining the rate at which recombination occurs. 
It was shown that there are strong
distortions of the CMB spectrum due to two-photon decay, but
unfortunately
these distortions  are practically unobservable because they lie in
the distant Wien region of the CMB spectrum.
\citet[hereafter P68]{1968ApJ...153....1P}, and 
\citet[hereafter SSS99 and SSS00]{1999ApJ...523L...1S,2000ApJS..128..407S} 
came to the same conclusion.
\cite{1975SvAL....1..196D} proposed to look 
for CMB spectral distortions due to transitions between higher
levels (Balmer, Paschen and higher series), 
which should be broadened due to the delayed 
recombination, and many papers were devoted to the computation of
these very small but extremely interesting effects 
\citep{1991Ap.....34..124G,1993obco.symp..548R,2004AstL...30..509D}.
Although several computations of the intensity of these
lines gave contradictory results (see e.g. \cite{1993obco.symp..548R} and 
\cite{2004AstL...30..509D}), it is obvious
that the corresponding distortions of the CMB spectrum are extremely weak,
and that $\Delta I/B_\nu$ does not exceeds the level of $10^{-7}$, 
which is two orders of magnitude below the sensitivity of COBE-FIRAS
experiment, but certainly will be the goal of future more precise
experiments. 

Another physical process with information about
the epoch of recombination is Rayleigh scattering on neutral hydrogen
and helium atoms \citep{2001ApJ...558...23Y}.
This process, due to the characteristic dependence of the scattering
cross-section on wavelength ($\sigma_R \propto \nu^4$), 
has a very strong frequency dependence of the power spectrum,
$C_\ell$. 
Unfortunately, this frequency dependence is rather similar to
the spectrum of dust emission from local dust and bright extragalactic 
star-forming galaxies, and the amplitude of the effect becomes
important in frequencies where the dust emission is significant
($\nu \ga 200$~GHz). 
This will make a rather difficult task to separate it from nearby
and bright foreground dust contribution.

In this paper, we concentrate on the detection of
hydrogen lines, and we propose a different method to 
detect the consequences of the overpopulated levels
of hydrogen atom at the epoch of recombination. 
We propose to look for frequency dependent angular fluctuations
of the CMB in different angular scales and at frequencies in the 
vicinity of the redshifted Balmer (and higher series) lines.
Here, we estimate how the small optical depths connected with these
transitions influences the angular distribution of the CMB,
arising mainly due to Thomson scattering on free electrons.
We show that one should
expect changes in the angular power spectrum due to these lines 
of about $0.3$~$\mu K$, being maximum 
at small angular scales ($\ell\approx 870$).
These numbers are at the level of the planned sensitivity of the Planck
satellite but for large angular scales, 
and certainly should be achieved by future
ground-based instruments looking for polarization of the CMB, and  
future space missions like CMBPol. It is important to 
mention that present day technology permits one to achieve sensitivities
at least 50 times better than that of Planck \citep{cmbpol}. 
The maximum signal that we are discussing corresponds to the 
redshifted H$\alpha$ recombination line, which unfortunately is
located in a frequency region where the contribution 
from dust emission is very important, and where the Rayleigh scattering
is 100 times larger in $C_\ell$ (10 times in temperature). 

\begin{figure}
\includegraphics[width=\columnwidth]{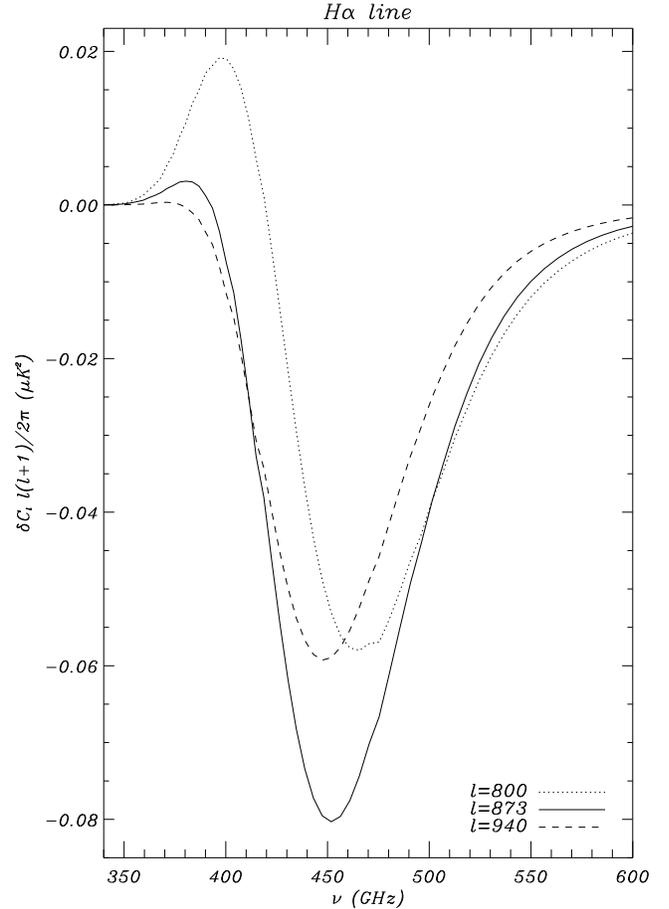}
\caption{Distortion of the power spectrum due to coherent 
scattering in the H$\alpha$ line, as a function of the observing
frequency today. The maximum distortion 
(in absolute value) is obtained at $\nu=452$~GHz for $\ell=873$.
We show that shifting the observing multipole in a region of 
$\Delta \ell = \pm 70$ around the maximum, we still have the signal
within the 70\% of the peak intensity. }
\label{fig:intro1}
\end{figure}

\begin{figure}
\includegraphics[width=\columnwidth]{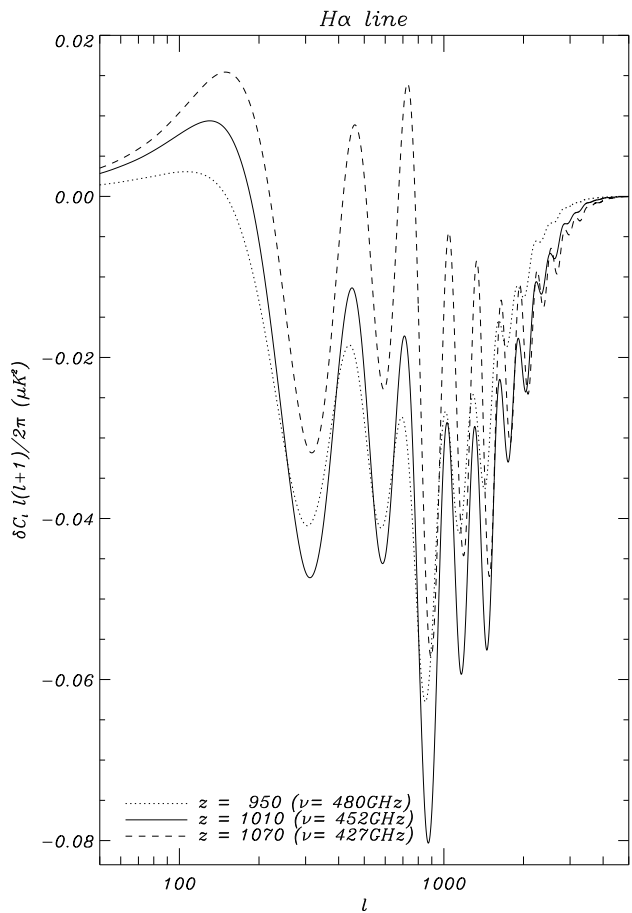}
\caption{Distortion of the power spectrum due to scattering in the
H$\alpha$ line as a function of angular
scale. }
\label{fig:intro2}
\end{figure}

Coherent scattering in the H$\alpha$ line produces a  
$\Delta T/T_0 \approx 10^{-7}$ signal in the angular distribution of
the CMB, whereas spectral distortions in the same line are
close to $10^{-8} - 10^{-7}$. Currently, 
experimentalists put their main effort in observing angular
fluctuations.
Therefore, the signal under discussion in this paper might be
observed sooner or later. 

This effect is small, but it leads to relatively narrow spectral
features in the power spectrum ($\Delta \nu/\nu \sim 10^{-1}$).
In addition, there are several regions in the angular power spectrum
at high multipoles where the sign
of $\delta C_\ell$ (=$C_\ell^{\rm line} - C_\ell$) 
connected with this process depends on the
wavelength, i.e. it changes from positive to negative when the 
frequency is shifted from 350~GHz to 450~GHz (Fig.~\ref{fig:intro1}).
The effect also has a very peculiar dependence on $\ell$ at all 
frequencies (Fig.~\ref{fig:intro2}), 
and these peculiarities in frequency and angular scale 
dependence might permit future observers to separate coherent scattering on
$H\alpha$ during the epoch of recombination from foregrounds and even 
from the cosmic variance of the effect of Rayleigh
scattering.

The discovery of these spectral features in the CMB angular
distribution will permit us to measure directly the tail of the
recombination process and its position with good 
precision, and may give additional information about cosmological
parameters of the Universe and their evolution with time.

This may also give direct
spectroscopic evidence of recombination. 
If Rayleigh scattering probes the existence of neutral hydrogen
immediately after the peak of the 
visibility function, the profile of the Balmer
lines features in the power spectrum gives us the possibility to check the
dynamics of recombination when the electron density changes from
$n_e/n_H\approx 1$ to $10^{-2}$, and correspondingly, 
the optical depth in H$\alpha$ and the population of the 
second level are traced. 
The characteristic physical length producing fluctuations in H$\alpha$ in that
epoch corresponds to smaller scales than those for Rayleigh
scattering. 
It is also important that for the hydrogen lines, 
each observed frequency corresponds to a given 
redshift $z$, but the 'picture' from Rayleigh scattering is the same
for all frequencies (there is only a change in amplitude). 
Thus, when the line is placed inside recombination, the
shape of the angular power spectrum for these two effects differs
strongly. 

Computations of the $\delta C_\ell$ connected with recombinational
lines are direct, because detailed estimates of the
recombination process show (SSS99)
that the assumption of ZKS68 about the Saha equilibrium
of all levels above level two is valid with good precision.
This gives us the possibility to find the optical depth in
each line of interest, and include it as an additional
opacity coefficient in the CMBFast code \citep{cmbfast}, as it was done
for the lines in the case of resonant scattering
by \cite{2002ApJ...564...52Z} (lithium doublet), and 
\cite[hereafter BHMS04]{basu04} for
fine structure lines of neutral atoms and ions of carbon, nitrogen 
and oxygen. 
Let us mention here that the computations of the line 
intensities (i.e. spectral distortions of CMB) require the
detailed calculation of tiny deviations from Saha equilibrium,
which are connected to the process of recombination 
controlled by 2s-level decay.
For our purposes, it is enough to use the equilibrium distribution
because the generation of new angular fluctuations 
is connected with just re-distribution of radiation over angles, so 
it has no relation to the difference between excitation
temperature and color temperature of radiation.
Optical depths in all lines, including Balmer lines, is very small
(e.g. $\tau \sim 5\times 10^{-5}$ for H$\alpha$ line at $z=1000$). 
But as it was shown by \citet[hereafter HMS05]{chm} and BHMS04, 
correlation effects with the existing 
radiation fluctuation field amplifies the effect
in such a way that 
$\delta C_\ell$s are effectively proportional to $\tau_\nu$, instead 
of $\tau_\nu^2$.
It should be noted that the signal that we are discussing is smaller
than the cosmic variance level, but as is explained by HMS05 and BHMS04, 
multi-frequency observations permit one to avoid the limit imposed by the
cosmic variance associated with the intrinsic CMB fluctuations, and
reveal signals below this threshold.

Finally, it is important to mention that these detections 
strongly rely on a accurate cross-calibration between 
different channels of a given experiment. 
However, there is a possibility to use 
the thermal SZ effect \citep{sunyaev72} on 
clusters of galaxies and sum of blackbodies approach for such a
calibration. Both methods are discussed by \citet{2004A&A...424..389C}.

The outline of the paper is as follows. In Section 2, we review the 
basic equations which describe the cosmic recombination process, and
we present a code that we have developed to 
compute the optical depths in the lines for the required transitions. 
In Section 3, we show how to derive the $\delta C_\ell$ quantities
for a given line, and present our results for the considered transitions. 
Section 4 presents a discussion about the foreground contamination and
the amplitude of the effect compared to Rayleigh scattering. 
Conclusions are given in Section 5.

\section{Cosmological recombination process}

In this section, we review the basic equations 
describing the process of cosmic recombination of hydrogen and helium,
and we show how to use them to infer the optical depths for
the recombination lines.
Since the first computations by ZKS68 and P68,  
many refinements have been introduced 
(for a review, see SSS00). 
SSS00 has the most detailed calculations to date,
including $300$ levels for the hydrogen atom, $200$ for HeI and $100$ for
HeII. 
When compared with the standard ``effective three-level computation'', 
their calculation results in a roughly 10\% change in the electron
fraction at low redshift for most of the cosmological models, plus a
delay in HeI recombination. All these results were incorporated in
the RECFAST code (SSS99), which consists in a 
modification of the effective three-level model to reproduce
the new values. 

For this paper, we are interested in computing 
the optical depths associated with the hydrogen 
recombination lines of the Balmer, Paschen and Brackett series. 
We will not discuss here the helium lines
(see \cite{1997AstL...23..565D} for a discussion on 
the spectrum of primordial HeI and HeII recombination lines). 
In order to obtain the optical depths, we need to know the 
populations for all levels in the hydrogen atom which contribute
to the transitions we want to study.
We will do this following two different approaches. 

On the one hand (method I), we will follow SSS00 and we will evolve the level
populations for a hydrogen atom with $10$ levels, separating levels
both by principal quantum number $n$ and angular momentum $l$. 
In method II, we will adopt the ionisation fraction from 
RECFAST and compute, using the Saha equation relative to the continuum, 
the population of the excited levels (e.g. \citet{1983A&A...123..171L}). 
This method is a valid approximation for those 
levels above the second one and for redshifts $z\ga900$, but 
as we will see, this is indeed our range of interest because at lower 
redshifts the visibility function is very small.

\subsection{Method I: Detailed follow-up of level population}

The detailed formalism and the equations 
describing the evolution of the population of the
levels of hydrogen and helium atoms with cosmic
time is presented in SSS00.
Here, we enumerate the assumptions of our work and the differences
to that paper.
\newcounter{paquito}
\begin{list}{\alph{paquito})}{\usecounter{paquito}\setlength{\rightmargin}{\leftmargin}}
\item We follow in detail all sub-levels of the hydrogen atom, taking
into account the quantum number $l$ for angular momentum, and the
principal quantum number $n$  
up to a maximum value of $n_{max}=10$. Thus, each state is labeled
with two numbers, $(n,l)$, and we have a total number of
$N=n_{max}(n_{max}+1)/2=55$ bound levels. It should be noted that SSS00 
do not separate different levels according to the quantum number $l$, 
but this is included in the computation of \cite{1993obco.symp..548R}.
We decided to follow all separate $l$ levels to cross-check our results
for the intensity of spectral lines with those from
\citet{1993obco.symp..548R} and \citet{2004AstL...30..509D}. 
\item Only radiative rates will be taken into account when solving
the set of equations, because collisional processes are not important
in the early Universe (see ZKS68, P68 and SSS00).
\item For helium, we will not make a detailed follow-up of all
levels, and we only evolve the equation for its total ionization
fraction (see RECFAST code, SSS99).
\item It will be assumed that the radiation field is a perfect
blackbody (COBE/FIRAS showed that deviations from blackbody are 
smaller than $10^{-4}$ \citep{1994ApJ...420..439M,1996ApJ...473..576F}), so 
the equation for the evolution of the radiation field will be
omitted. This approximation is not valid in the case of the Lyman
series for hydrogen, where the spectral distortions are strong, 
but it is good for Balmer series and above.
\item We will use the Sobolev escape probability method 
\citep{sobolev} to deal with the
evolution of resonance lines. This will decouple the evolution of
the radiation field from the evolution of the level population.
We note that this method was developed
for a completely different problem. For a derivation of this
method in a cosmological situation, see \citet{2004AstL...30..657D}.
\end{list}

Values for the physical constants for hydrogen and helium 
were taken from SSS99 and SSS00. 
The values for the oscillator strengths and the corresponding 
Einstein coefficients for instantaneous emission were computed following 
\citet{1957ApJS....3...37G}. 
The value for the two photon decay transition $A_{2s,1s}$ 
is essential, given that
this is the dominant mechanism driving the
cosmological recombination. We adopt here the latest 
value, $8.22458 s^{-1}$, from \cite{1989PhRvA..40.1185G}. 

The photoionization/recombination rates were computed in
two different ways.
The first one was by using the photoionization 
cross-sections from 
\cite{1991CoPhC..66..129S,1994MNRAS.268..109H}.
The second one uses the analytical expressions from 
\cite{1958MNRAS.118..477B},
which are valid for small values of the energy
of the ejected electron.
In both cases, radiative recombination rates 
included spontaneous and induced transitions.  
Our final results were obtained 
using the second method, which is computationally faster than
the first one. Nevertheless, we checked that both methods 
give similar results for the populations of the levels.

Once the populations of all levels are obtained at all redshifts, 
the optical depth associated with any permitted 
transition $j \rightarrow i$ connecting the levels $i=(n,l)$ and
$j=(m,l')$ (with $m>n$) can be computed using the Sobolev formula as
\begin{equation}
\tau_{ij} = 
\frac{A_{ji} \lambda_{ij}^3 [n_i (g_i/g_j)-n_j]}{8\pi H(z)}
\label{eq:tau}
\end{equation}
where $A_{ji}$ is the Einstein coefficient for instantaneous emission, 
$g_i$ is the degeneracy of the energy level, and $H(z)$ is the 
(time dependent) Hubble constant. 
The total optical depth $\tau_{nm}$ 
associated with a particular line $m\rightarrow n$
is easily derived by adding the contribution from all the 
permitted transitions connecting the sublevels with those 
principal quantum numbers.

\begin{figure}
\includegraphics[width=\columnwidth]{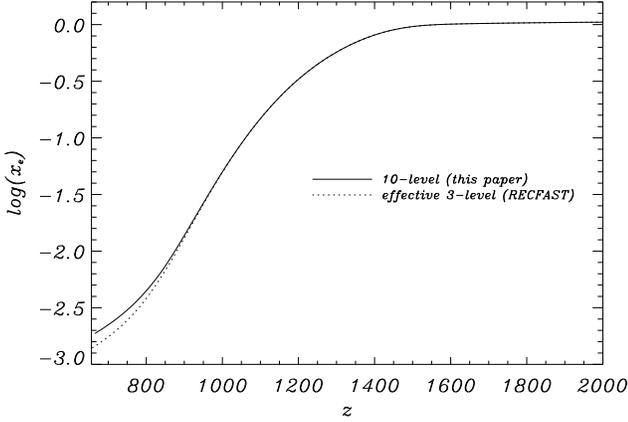}
\caption{Multilevel hydrogen recombination as computed in this paper
(using 10 levels for quantum number $n$, and 
following independently all the states 
of angular momentum), compared with the standard 'effective three level'
computation from RECFAST. It is shown $x_e$($=n_e/n_H$) as a function of
redshift in the vicinity of the epoch of cosmic recombination for the
cosmological model with parameters $\Omega_b=0.044$, $\Omega_{tot}=1$
and $h=0.71$ \citep{2003ApJS..148....1B}. }
\label{fig:x_e}
\end{figure}

In Fig.~\ref{fig:x_e} we show our result for the
electron density evolution ($x_e = n_e/n_H$) as a function of
redshift
\footnote{Note that in our notation, $n_H=n_0 (1+z)^3$ is 
the total number density of hydrogen, so $n_H = n_{HI} + n_p$.}, 
compared with the standard calculation using RECFAST. These
results are consistent with those obtained by SSS00 
when they use $n=10$ levels, 
but it should be noted that in our calculation 
we do a detailed follow-up of the
population of the different angular momentum states.
When more levels are included in the computation, SSS00 show that the 
residual electron density at low redshifts becomes smaller, 
and for $n\ga 50$ it 
converges as the atom becomes complete in terms of energy levels. 
However, for the purposes of this paper, it is enough to consider 
a 10-level hydrogen atom to achieve good precision in the region of interest.

\subsection{Method II: Saha equation from the continuum}

As suggested by ZKS68, and confirmed with the computations of SSS00, 
the approximation of equilibrium of all levels above the second one
yields very good results, and deviations greater than 10\% are only
found at redshifts $z \la 800$. 
Thus, we can derive (to a good approximation) the population of the
level in two steps. We first use the standard RECFAST computation, and
derive the evolution of the electron density. From here, and using the 
Saha equation relative to the continuum, we can derive
the population of all levels above the second one, and 
can compute all the optical depths for Balmer lines and
higher series. It should be noted that this approximation 
fails for redshifts below $z\sim900$ because most of the atoms
have recombined and the transition rates are not enough to keep
equilibrium between high levels.

Thus, the population of the level $n$ is derived in this second method
as 
\begin{equation}
n_i = \frac{n_e n_p h^3}{(2\pi m_e k_B T_e)^{3/2}} \frac{g_i}{2}
\exp(\chi_i/k_B T_e)
\label{saha}
\end{equation}
where the state $i=(n,l)$ has excitation energy below the continuum 
$\chi_i=13.598/n^2$~eV, and $g_i$ is the statistical weight of the
state. From that expression, we can derive the absorption 
coefficient for a transition $j\rightarrow i$, and from here and
using the Sobolev approximation, the optical depth for each line 
at redshift $z$ (equation~\ref{eq:tau}).

\subsection{Optical depth in the hydrogen recombination lines}

Although we have computed many more transitions, in this paper
we shall concentrate on those lines whose expected contribution
is largest. Thus, we investigate here the 
three first transitions (labelled as $m\rightarrow n$) for the Balmer 
($n=2$), Paschen ($n=3$) and Brackett ($n=4$) series. 
Table~\ref{tab:dataset} presents the wavelengths and oscillator strengths 
for these lines. 
For illustration, we show in the last two columns the
redshifted frequency and the  
optical depth in the line evaluated at redshift $z=1000$, and
for a cosmological model with parameter values taken
from \cite{2003ApJS..148....1B}, 
i.e. $\Omega_{tot}=1$, $\Omega_b=0.044$ and reduced Hubble constant $h=0.71$. 
Throughout this paper we have used these values, unless otherwise stated.

\begin{table*}
\caption{List of the strongest hydrogen recombination lines studied in this
paper. We present the three strongest lines for the first three series 
of hydrogen atom between excited states: Balmer ($m\rightarrow 2$), 
Paschen ($m\rightarrow 3$) and Brackett ($m\rightarrow 4$). 
For each transition $m\rightarrow n$, we show its name, the rest
frame wavelength, and the oscillator strength (these values
have been taken from \citet{1957ApJS....3...37G}). 
For illustration, we also present in the fifth column the 
redshifted frequency observed today if emission
took place at $z=1000$. The last column shows the optical depth in 
the line at that redshift ($z=1000$), as computed using our code.}
\label{tab:dataset}
\centering
\begin{tabular}{@{}lccccc}
\hline 
\hline 
Transition & $m$ & $\lambda_{nm}$ ($\AA$) & $f_{nm}$ & $\nu_0(z=1000)$ (GHz)
& $\tau _{nm}(z=1000)$ \\ 
\hline
\multicolumn{6}{c}{Balmer lines ($n=2$)} \\
\hline
H$\alpha$ &  3 & 6562.8 &      0.6407 & 456.3 &  4.53$\times 10^{-5}$ \\
H$\beta$  &  4 & 4861.3 &      0.1193 & 616.1 &  6.14$\times 10^{-6}$ \\
H$\gamma$ &  5 & 4340.5 &      0.0447 & 690.0 &  2.09$\times 10^{-6}$ \\
\hline
\multicolumn{6}{c}{Paschen lines ($n=3$)} \\
\hline
P$\alpha$ & 4 & 18751.0 &   0.8421 &  159.7 &    1.17$\times 10^{-7}$ \\
P$\beta$  & 5 & 12818.1 &   0.1506 &  233.6 &    1.47$\times 10^{-8}$ \\
P$\gamma$ & 6 & 10938.1 &   0.0558 &  273.8 &    4.75$\times 10^{-9}$ \\
\hline
\multicolumn{6}{c}{Brackett lines ($n=4$)} \\
\hline
B$\alpha$ & 5 & 40512.0 &    1.0377 &   73.9 &   2.57$\times 10^{-8}$ \\
B$\beta$  & 6 & 26252.0 &    0.1793 &  114.1 &   3.42$\times 10^{-9}$ \\
B$\gamma$ & 7 & 21655.0 &    0.0655 &  138.3 &   1.09$\times 10^{-9}$ \\
\hline
\hline                                                              
\end{tabular}
\end{table*}

\begin{figure}
\includegraphics[width=\columnwidth]{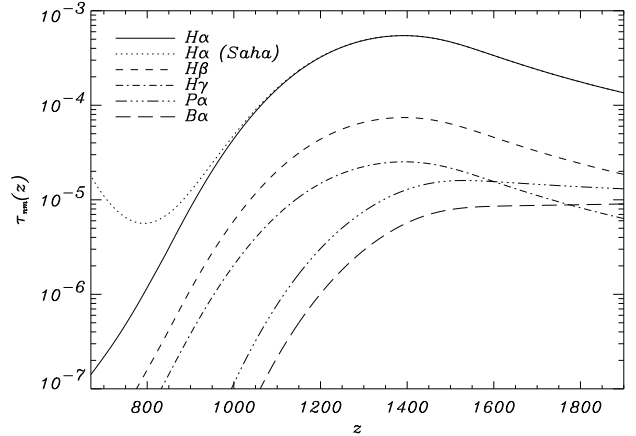}
\caption{Optical depths for several lines considered in this paper, as
a function of redshift $z$.
All values were computed using method I.  
For the H$\alpha$ line, it is also shown the computation using the Saha
approximation (method II), to show that with 
this method, deviations with respect to the 
detailed calculation (method I) only appear 
when the Universe is practically transparent, which
corresponds to the high frequency wing of the line and where the 
expected effect is not important. }
\label{fig:taus}
\end{figure}

Figure~\ref{fig:taus} shows the optical depth in the lines 
($\tau_{nm}$) derived with our method I, and for this cosmological model. 
Using the method II, we obtain these same values for the optical 
depths, but only for redshifts above $z\approx900$.
To illustrate this point, we also display in Fig.~\ref{fig:taus} 
the optical depth for the H$\alpha$ line using the method II (dotted line).
Method II reproduces 
the non-equilibrium computation above $z\approx900$, but fails below this
point, showing a divergent behaviour. The reason for this 
was pointed out above. At these redshifts, the radiation field 
has not enough photons to maintain the equilibrium of the second
level with the continuum. The photoionization rates become smaller than 
the photorecombination rates, and the Saha equilibrium between the second
level and the continuum is no longer valid (e.g. ZKS68). 
Moreover, the populations of  
$2s$ and $2p$ levels start to show strong departures 
from their equilibrium ratio.
Given that the cosmological recombination is proceeding much more slowly
than expected from a Saha recombination, 
method II (which uses the Saha equation relative to a continuum 
level which has been computed using the effective
three level calculation) predicts very high values for the 
population of the second level in this redshift range. 
This is the reason for the divergent behaviour at $z\la 900$
of the H$\alpha$ optical depth computed with method II. 
Although we have used the exact (non-equilibrium) computation for all the
results presented in this paper, we wanted to show that using a very simple
approximation (method II) it is possible to correctly estimate the 
values for the optical depths in the range of interest (as we show below, 
the peak of the effect we are investigating occurs at $z\sim 1010$). 

In Fig.~\ref{fig:taus}, we finally note that 
for Balmer transitions, the shape of the 
optical depth as a function of frequency is similar for all lines.
This can be easily understood 
because the population of the 
second level is much larger than the population of higher levels, and
thus the optical depth is directly proportional to this population 
($n_{2s}+n_{2p}$) for all Balmer lines.

\begin{figure}
\includegraphics[width=\columnwidth]{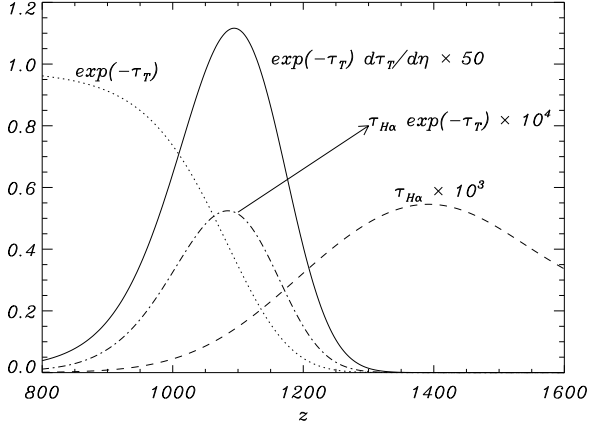}
\caption{Dependence of the optical depth with the redshift for the H$\alpha$
  line. For comparison, we also show 
the Thomson visibility function ($\mathcal{V}_T(\eta) = 
(d\tau_T/d\eta) \exp(-\tau_T)$),  
the function $exp(-\tau_{T})$, and the function 
$\tau_{H\alpha} exp(-\tau_{T})$. To display all of them on the same scale,
they were rescaled by the indicated factors. }
\label{fig:visfunc}
\end{figure}

Figure~\ref{fig:visfunc} shows the optical depth in the
H$\alpha$ line, together
with the Thomson visibility function \citep{SZ70}, 
$ \mathcal{V}_T(\eta) = (d\tau_T/d\eta) \exp(-\tau_T)$, and the
function $\exp(-\tau_T)$. Here, $\tau_T$ is the Thomson optical depth
and $\eta$ is the conformal time. As we can see from this figure, 
the maximum optical depth for this transition is reached beyond the
peak of the visibility function, around $z\approx1400$, but there 
the Thomson optical depth is very large. 
This is why the maximum contribution to the
effect we are discussing comes from lower redshifts ($z\sim 1000$).  
As we show in the next section, the generation of new anisotropies is conected
to the term $\tau_{H\alpha} \exp(-\tau_T)$, also shown in
Fig.~\ref{fig:visfunc}.

\section{Imprint of hydrogen recombination lines on the CMB}

The imprint of line transitions on the CMB spectrum has been
examined by \cite{2002ApJ...564...52Z} for the
case of lithium recombination, and for other ions and atoms in
BHMS04. We extend these works here and we consider the case
in which the coherent scattering occurs inside recombination.
 
The drag force induced on CMB photons 
by the scattering in the hydrogen recombination lines was already
discussed by \cite{2001ApJ...555L...1L}. They showed that due to the low
population of the excited levels, 
the characteristic time over which the peculiar velocity of
the gas is damped due to the drag force on the hydrogen
atoms is much longer than the Hubble time 
at that epoch. We repeated this computation and found that it is at
least five orders of magnitude longer.
Thus the drag force can be safely neglected when computing
the effect of coherent scattering in these lines on the power spectrum
of the CMB, as it was done in \cite{2002ApJ...564...52Z} and BHMS04.

\subsection{Widths of the recombination lines}

Before hydrogen is significantly recombined in the Universe, 
the optical depth for 
electron scattering is very high, so every scattering leads to a 
broadening of the line. For $z=1000$, the thermal (Doppler) width of
the lines is close to $(\Delta \nu/\nu)_{th} \approx 4\times10^{-5}$. 
Subsequent electron scattering of these photons might increase this
width up to $(\Delta \nu/\nu)_{e} \approx 2\times10^{-3}$. 
The characteristic width of recombination in redshift space, 
$(\delta z/z)_{\rm rec}$, is
much bigger than this electron broadening (from Fig.~\ref{fig:visfunc},
the visibility function has a width of $\delta z \approx 80$, so 
$(\delta z/z)_{\rm rec} \approx 1/14$).
On the other hand,  the optical depths in the lines
under discussion are also changing slowly. The redshift 
interval $(\delta z/z)_{H\alpha}$ in which $\tau_{H\alpha}$ 
changes significantly (i.e. $\delta \tau_{H\alpha}/\tau_{H\alpha}
\approx 1$) is about 0.07 for $z=1000$.
Thus, we expect that both Doppler and 
electron broadening of the lines will not change our effect 
significantly. 
We have checked this point by repeating the computations of this
paper for different widths for the lines. 

The opacity in each line ($m \rightarrow n$) is computed as 
\begin{equation}
\dot{\tau}_L \equiv \frac{d\tau_L}{d\eta} = \tau_{nm} \; 
\mathcal{P}(\eta; \eta_L)
\label{eq:tau_l}
\end{equation}
where $\mathcal{P}(\eta; \eta_L)$ is a profile function of area unity, and
centered at $\eta_L$, the conformal time corresponding to
the redshift that we are observing ($1+z_{obs} =
\nu_{nm}/\nu_{obs}$). 
For simplicity, we have adopted here a normalized Gaussian profile, so 
\[
\mathcal{P}(\eta; \eta_L) = 
\frac{\exp( -\frac{(\eta-\eta_L)^2}{2\sigma_L^2} )}
{\sqrt{2\pi \sigma_L^2}} 
\]
where $\sigma_L$ is the width of the Gaussian. 
We have explored three different values for
$\sigma_L/\eta_L$, namely $10^{-3}$, $5 \times 10^{-3}$ and $10^{-2}$,
but we found similar results for all of them, as expected. 
All values quoted in this paper correspond to the 
case $\sigma_L/\eta_L=0.005$.

One important consequence that can be determined from here is that every
frequency observed today corresponds to a given redshift $z$. 
With electron scattering, we are averaging all the effects well before
the peak of the visibility function, and we are losing information about the
redshift in which a given part of the signal was produced.
However, the study of these lines will permit us to check 
the velocities and optical depths at any stage of
recombination. 
In order to do this, 
it will be necessary to optimise the observing frequencies
of the detectors, and to have observing bandwidths narrower than 
the widths of the features (we will see below that
typical widths are $\Delta \nu/\nu \approx 0.1$). 
Unfortunately, present day and planned experiments (like Planck)
have widths of the channels broader than this, 
and even broader than the width of recombination, 
so these effects are averaged inside the observing bandwidths.  
But in principle, future ground based experiments or experiments like
CMBPol might be adapted to the width and
frequency dependence of the features that we are discussing, and 
 it could be possible to trace using this
{\it tomography} the overall behavior of recombination.

\subsection{Coherent scattering in hydrogen lines during recombination}

In this subsection, we present the equations and the method of
computation of the effect of coherent scattering in hydrogen lines.
To perform the computations for this paper, we have used the code from
BHMS04, which was a modification of CMBFAST including 
the presence of a resonant line at a given redshift. We will 
follow here their notation.

In the conformal Newtonian gauge, the Boltzmann equation for the
evolution of the $k$-mode of the photon
temperature fluctuation can be formally integrated to give
\citep{cmbfast}
\begin{equation}
\Delta_T(k,\eta_0,\mu) = \int_{0}^{\eta_0} d\eta 
e^{i k \mu (\eta - \eta_0)} e^{-\tau} \times 
\Bigg[ \dot{\tau} (\Delta_{T0} + i\mu v_b) 
+ \dot{\phi} - i k \mu \psi \Bigg]
\label{eq:deltaT}
\end{equation}
where we have explicitly dropped the polarization term, which
contributes at most a few percent of the total signal and will be
neglected here. $\Delta_{T0}$ accounts for the intrinsic fluctuations; 
$v_b$ is the velocity of baryons; $\phi$ and $\psi$ are the scalar
perturbations of the metric in this gauge, and 
$\mu = \mathbf{\hat{k} \cdot \hat{n}}$. 
The total optical depth is defined here as 
$\tau(\eta) = \int_{\eta}^{\eta_0} \dot{\tau} d\eta$, 
and for our problem, 
it has contributions from both Thomson and coherent scattering. 
Thus, using equation \ref{eq:tau_l}, we have that for a 
given line $L$, the total optical depth is 
$\dot{\tau} = \dot{\tau}_T + \dot{\tau}_{L} = a n_e(\eta) \sigma_T + 
\tau_{nm} \mathcal{P}(\eta;\eta_L)$. 

Using this modification inside the CMBFAST code, and neglecting the drag
force induced by this process on the CMB photons,  
it is possible to compute the effect of the coherent
scattering on the power spectrum. 
The important point for us is that, as shown
by HMS05, the effect of lines on the CMB spectrum is
amplified due to correlation with the existing radiation field.
The presence of coherent scattering in lines produces an angular
redistribution of the CMB photons. For the lines considered in
BHMS04, the net effect is generation of anisotropies at large
scales due to Doppler motion of scattering atoms, and suppression of
power at small scales. 
Summarizing, the amplification effect consists in that 
the change in the power spectrum ($\delta C_\ell \equiv 
C_\ell^{line} -C_\ell$) due to a given transition $m \rightarrow n$ 
is proportional to $\tau_{nm}$ (and not to $\tau_{nm}^2$) for small values
of the optical depths. Thus, in the optically thin limit we have
\begin{equation}
\delta C_\ell = \tau_{nm} \cdot \mathcal{C}_1(\ell) + 
\tau_{nm}^2 \cdot \mathcal{C}_2(\ell) + 
\mathcal{O}(\tau_{nm}^3)
\label{delta_Cl}
\end{equation}
were the equation for the functions $\mathcal{C}_1(\ell)$ and 
$\mathcal{C}_2(\ell)$ is given in Appendix A of BHMS04 (note that 
in that equation, the polarization terms were neglected). 
In the limit of  high-$\ell$ values, they obtained  the well-known
result of $\mathcal{C}_1(\ell) \rightarrow -2 C_\ell$. 

\begin{figure}
\includegraphics[width=\columnwidth]{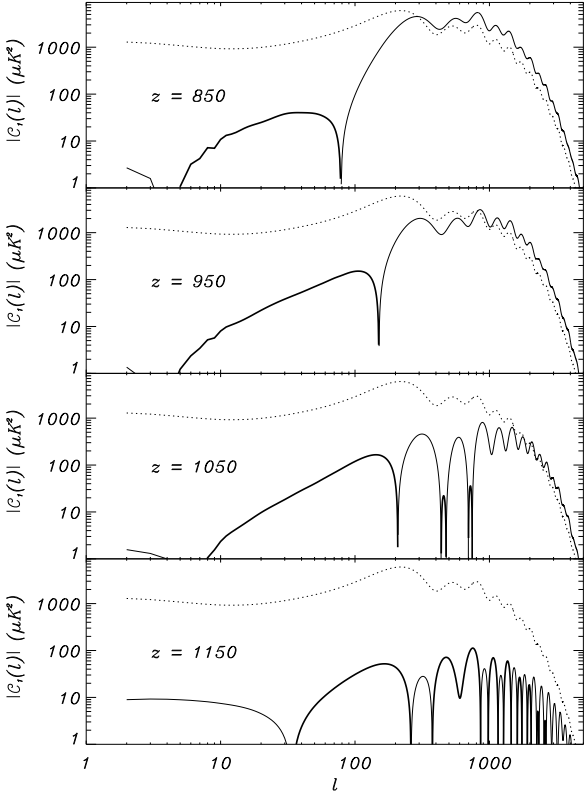}
\caption{Linear term of the correction to the angular power spectrum
originated from coherent scattering in a line located 
at the specified redshift.
These curves have to be multiplied by the optical depth in the line to
obtain the first order correction to $\delta C_\ell$. 
Thick lines correspond to positive values, and
thin lines to negative ones. For comparison, it is also shown the 
primary CMB power spectrum (dotted lines).
Note that at high-$\ell$, the correction is roughly $-2 C_\ell$ for the case
$z=850$ or $z=950$, as 
we would expect for a line located between us and recombination.
This property disappears when we go to higher redshifts, because in that
case the line is located inside recombination where radiative viscosity and
thermal conductivity are important. [See movie in the electronic
version]. }
\label{fig:linear_term}
\end{figure}

Eq.~\ref{delta_Cl} can be used in our problem, in which the scattering line
is embedded inside recombination, although there are some differences
to the case in which the line is placed between recombination and us. 
As was shown in BHMS04, the generation and blurring
terms tend to cancel each other at a multipole value
$\ell \sim (\eta_0-\eta_{cs})/(\eta_{cs}-\eta_{rec})$, with 
$\eta_{cs}$, $\eta_{dec}$ and $\eta_0$ the conformal times at the
coherent scattering, at recombination and the present day, respectively.
From this, one can expect that if coherent scattering takes place not 
far from recombination, both generation and blurring will cancel
over a wider range of multipoles. 
We illustrate this in Fig.~\ref{fig:linear_term}, where we show the
dependence of the linear term ($\mathcal{C}_1(\ell)$) for four
different cases of a hypothetical line placed at redshifts 
$z=850, 950, 1050$ and $1150$. 
Thus, in other words, the presence of this strong decrease in the power 
at angular scales $\ell \la 200$ is 
direct evidence that the line is 
located at the epoch of recombination in the Universe.  
Unfortunately, because of this decrease, the detection at large
angular scales of this signal will be more difficult.

Only at redshifts lower than the redshift of recombination is the
blurring term dominant at high multipoles: when the line 
is well within the peak 
of the Thomson visibility function, both blurring and generation terms 
cancel each other, at least to a level of a few percent.
This is seen in Fig.~\ref{fig:linear_term}, so for redshifts in the
tail of the visibility function ($z=850$) the linear term is equal 
to $-2 C_\ell$ at high multipoles. But if we move to higher redshifts,
this behavior disappears. 

\begin{figure}
\includegraphics[width=\columnwidth]{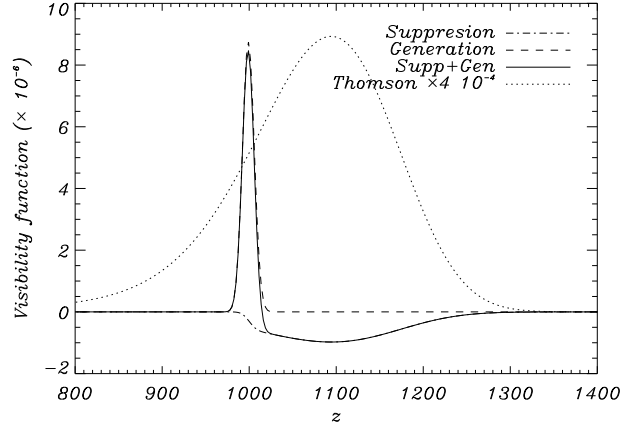}
\caption{Correction to the visibility function in the presence of
coherent scattering in the H$\alpha$ line. To first order in the
optical depth of the line ($\tau_{nm}$), there are two terms. One is
associated with suppression (eq.~\ref{eq:V_supp}),
and the other one with the generation of new anisotropies  
(eq.~\ref{eq:V_gen}). Both terms are plotted
(dot-dashed and dashed lines, respectively) together with 
their sum (solid line), for the case of observing the H$\alpha$ line
placed at $z=1000$. For comparison, we also show the  
Thomson visibility function (dotted line). }
\label{fig:visfunc2}
\end{figure}

For illustrational purposes, 
we discuss how the visibility function is modified when we 
consider coherent scattering in the hydrogen lines.
In the optically thin limit, we can write to first order in the
optical depth ($\tau_{nm}$) that 
\begin{equation}
\mathcal{V}(\eta ; \eta_L) = 
\dot{\tau} e^{\textstyle -\tau } \approx 
[1 - \tau_{nm} \mathcal{A}(\eta)] \mathcal{V}_T + 
\tau_{nm} \mathcal{P}(\eta) e^{\textstyle -\tau_T } 
+ \mathcal{O}(\tau_{nm}^2)
\end{equation}
where we have defined the area function of the profile,
$\mathcal{A}(\eta)$, as
\[ 
\mathcal{A}(\eta) = \int_{\eta}^{\eta_0} d\eta' \mathcal{P}(\eta')
\]
From here, we see that to first order in the optical
depth, we have two terms giving us the correction to the 
Thomson visibility function: one associated with suppression 
\begin{equation}
\mathcal{V}_{supp} = - \tau_{nm} \mathcal{A}(\eta) \mathcal{V}_{T} = 
 - \tau_{nm} \mathcal{A}(\eta) \dot{\tau}_T  e^{\textstyle -\tau_T}
\label{eq:V_supp}
\end{equation}
and another connected to generation of anisotropies:
\begin{equation}
\mathcal{V}_{gen} = \tau_{nm} \mathcal{P}(\eta) e^{\textstyle -\tau_T}
\label{eq:V_gen}
\end{equation}
In Fig.~\ref{fig:visfunc2} we display these two terms 
for a case in which we are observing the H$\alpha$ 
line placed at $z=1000$ (i.e. we are observing at $\nu_0=456$~GHz), and
with an instrumental width of $\sigma_L/\eta_L=0.005$ .  
The dot-dashed line corresponds to the suppression
term, whereas the dashed line shows the generation one. The former is
proportional to the Thomson visibility function for redshifts
{\em larger} than the redshift being probed by the line, since only
anisotropies located {\em behind} the line can be blurred. On the other
hand, the generation of anisotropies is located at the position of the
line, and its amplitude is weighted by the amount of free electrons
situated {\em between} the line and the observer (factor $\exp({-\tau_T})$
in eq.~\ref{eq:V_gen}).
The curve $\tau_{H\alpha} \exp(-\tau_T)$  was shown in 
Fig.~\ref{fig:visfunc}.

\subsection{Results for the considered lines}

\begin{table*}
\caption{Maximum signal (in absolute value) obtained for each of the
lines within the considered cosmological model.  
It is shown the angular multipole, the redshift and the observed
frequency (today) at which we have maximum effect. 
We also present the (full) widths $\Delta \ell$ and $\Delta \nu_0$ around
the peak value, defined as the regions where the signal is
greater than $70$\% of the maximum value. 
The last two columns show the amplitude of the effect in temperature 
($\mu K$) and the relative value of the temperature to the primordial
CMB power spectrum. All these maximum deviations are negative
(i.e. the correction to the power spectrum $\delta C_\ell$ at those
maxima is negative). }
\label{tab:maximos}
\centering
\begin{tabular}{@{}lccccccc} 
\hline
\hline
Line  & $\ell_{max}$ & $\Delta \ell $  & $z_{max}$  & $\nu_0$ &
$\Delta \nu_0$ &   
$\sqrt{ \ell (\ell+1)|\delta C_\ell| /2\pi}$ & 
$\sqrt{ |\delta C_\ell|/C_\ell}$ \\
 & & & & [GHz] & [GHz] & [$\mu K$] & \\
\hline
H$\alpha$   &           873 &  140&        1010 &           452 & 56 &      
0.28 &    $5.8\times 10^{-3}$ \\  
H$\beta$   &           873 &   140&       1010 &           610 &  76 &      
0.10 &    $2.1\times 10^{-3}$ \\  
H$\gamma$   &           873 &  140&        1010 &           683 & 85 &     
0.06 &    $1.2\times 10^{-3}$ \\  
P$\alpha$   &           888 &  121&        1050 &           152 & 16 &   
$1.6\times 10^{-2}$  &   $3.5\times 10^{-4}$ \\  
P$\beta$   &           888 &   121&       1050 &           223 &  24&   
$5.7\times 10^{-3}$ &   $1.2\times 10^{-4}$ \\  
P$\gamma$   &           888 &  121&        1050 &           261 & 28&    
$3.3\times 10^{-3}$ &   $7.1\times 10^{-5}$ \\  
B$\alpha$   &           891 &  116&        1060 &            70 & 7&    
$8.1\times 10^{-3}$   &   $1.8\times 10^{-4}$ \\  
B$\beta$   &           891 &   116&       1060 &           108 & 11&    
$2.9\times 10^{-3}$ &   $6.4\times 10^{-5}$ \\  
B$\gamma$   &           891 &  116&        1060 &           130 & 13&    
$1.7\times 10^{-3}$ &   $3.6\times 10^{-5}$ \\  
\hline
\hline
\end{tabular}
\end{table*}

Our main result for the nine transitions under discussion 
can be summarized in Table~\ref{tab:maximos},
where we present the angular scale and the 
redshift at which we obtain the maximum signal for each of the
lines.
From these values, we see that the best angular scale to look for this
effect is placed in the region of the third acoustic peak, and that
the maximum signal comes from redshifts $z\sim 1010-1060$. These
values lie close to the peak of the visibility function, but 
in the low-redshift tail.
We also present the full widths of the regions around the maxima where the
signal is within 70\% of the peak value, both in multipole ($\Delta \ell$) 
and frequency ($\Delta \nu_0$). 
We note that $\Delta \nu_0/\nu_0$ is of the order of 10\%, which is
hundreds of times larger than the Doppler broadening of the lines.
This will help in the detection of these features.
We also show that the width of the feature in multipole space is
of the order of 140, so a full sky coverage is not necessary to detect
the signal. 

The last two columns in Table ~\ref{tab:maximos} show the amplitude of 
the signals at these maxima, both in temperature and the fractional
change in temperature with respect to the intrinsic CMB power
spectrum.
Although the values for 
$\sqrt{|\delta C_\ell|/C_\ell}$ are small, the present day sensitivity for
the detection of angular fluctuations is better than that for spectral
measurements, so in principle these features should be easier to
detect than the distortions of the spectrum.
We now investigate in more detail the angular and frequency
dependences of this effect. We will concentrate here on the H$\alpha$
line, which gives the strongest signal. Predictions for the other 
transitions can be derived in a similar way.

\begin{figure}
\includegraphics[width=\columnwidth]{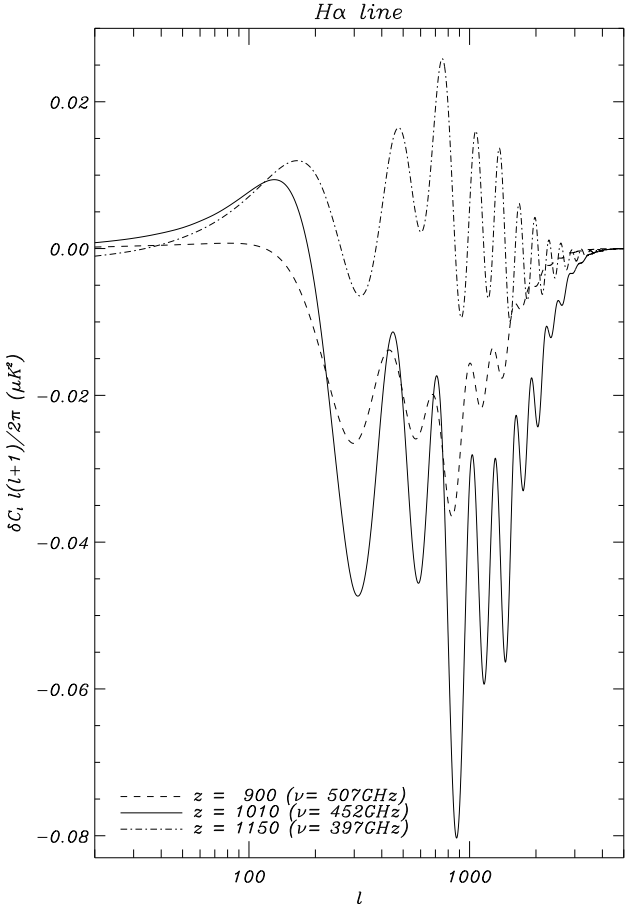}
\caption{Correction to the power spectrum due to coherent scattering
in the H$\alpha$ line. 
We plot three curves for three different redshifts, namely $z=900, 1150$ 
(which correspond to $507$~GHz and $397$~GHz, respectively), and also 
$z=1010$ ($452$~GHz), which is the redshift (frequency) at which we have 
the maximum amplitude for signal, 
at multipole $\ell=873$. Note how the $z=900$ case mimics 
the shape of the power spectrum at high multipoles, but with negative
sign, and how this behavior disappears as we go to higher redshifts. 
Observing at different frequencies with sufficiently narrow spectral 
bands, we can perform tomography of the recombination 
epoch using the H$\alpha$ line. }
\label{fig:features2}
\end{figure}

Fig.~\ref{fig:features2} shows the angular dependence of the effect, for the
case of H$\alpha$ line, and for redshift values $z=900$, $1010$ and
$1150$, which correspond to (redshifted) present day frequencies
of $507$, $452$ and $397$~GHz, respectively. 
Given the small values of the optical depth, 
the linear term gives the dominant contribution to the effect, 
so the shape of the plot is similar to Fig.~\ref{fig:linear_term}, 
but the amplitude is modulated by the
optical depth for the transition at every redshift. 
As shown in Table~\ref{tab:maximos}, the maximum signal is obtained 
at redshift $z=1010$. Prior to this redshift, the shape of the $\delta
C_\ell$ mimics that of the primordial $C_\ell$ at high multipoles, 
because the suppression term dominates here 
(and hence the $\delta C_\ell$ are negative). 
When the line is placed at higher redshifts, well inside the 
recombination, the cancellation of the generation and 
blurring terms at high multipoles
produces a number of regions where the $\delta C_\ell$ quantities
have different signs. The exact position of the zeros strongly depends on the
particular cosmological model, and as pointed above, it could be used
to investigate in detail the shape of the transfer function at all 
redshifts. 

\begin{figure}
\includegraphics[width=\columnwidth]{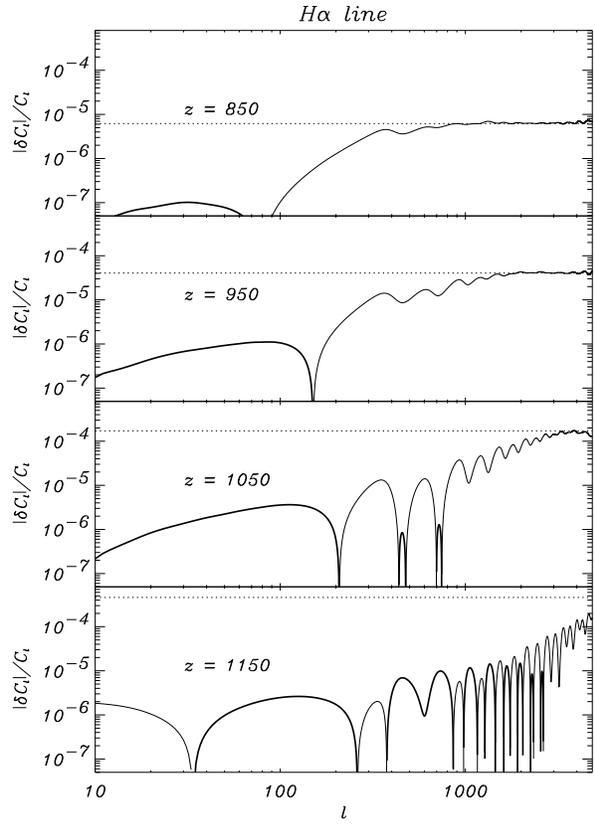}
\caption{Relative correction ($|\delta C_\ell| / C_\ell$) to the power
spectrum due to scattering in the H$\alpha$ line, as a function of
angular scale ($\ell$). Thick lines correspond to positive values
whereas thin lines represent negative ones. 
The distortions are larger at small angular
scales, approaching the asymptotics of $-2\tau_{H\alpha}$ (dotted line) for
redshifts before the peak of the visibility function. 
The region where this asymptotic
behavior is valid changes with redshift, and for redshifts well inside 
the recombination, it disappears. 
This behavior could give us information about the process of
dissipation of short wavelength adiabatic perturbations due to
radiative viscosity and thermal conductivity during recombination. }
\label{fig:deltaCl_to_cl}
\end{figure}

In Fig.~\ref{fig:deltaCl_to_cl} we present the same redshift slices as in
Fig.~\ref{fig:linear_term}, but now showing $|\delta
C_\ell|/C_\ell$ for the H$\alpha$ transition. 
The relative value of the distortion of the power spectrum is of the 
order of $10^{-5} - 10^{-4}$ in band power ($\sim 10^{-2}$ in
temperature). 
It should be also noted that 
$|\delta C_\ell|/C_\ell$ is growing at large multipoles, and
approaches the asymptotic value of $-2\tau_{H\alpha}$ (shown here as a
dotted line), although the multipole region where we reach this
behavior depends on the considered redshift slice.
When the line is placed at higher redshifts inside recombination, the 
asymptotic behavior is reached at higher multipoles  
in the damping tail region of the spectrum. 
Moreover, when the redshift is high enough, this asymptotic region 
disappears, and we find a complicated structure 
of positive/negative features due to the aforementioned cancellation. 
Thus, this method could give us information about the 'Silk damping'
\citep{1968ApJ...151..459S}, or 
processes of dissipation of short wavelength adiabatic 
perturbations due to radiative viscosity
and thermal conductivity during the recombination.

\begin{figure}
\includegraphics[width=0.8\columnwidth]{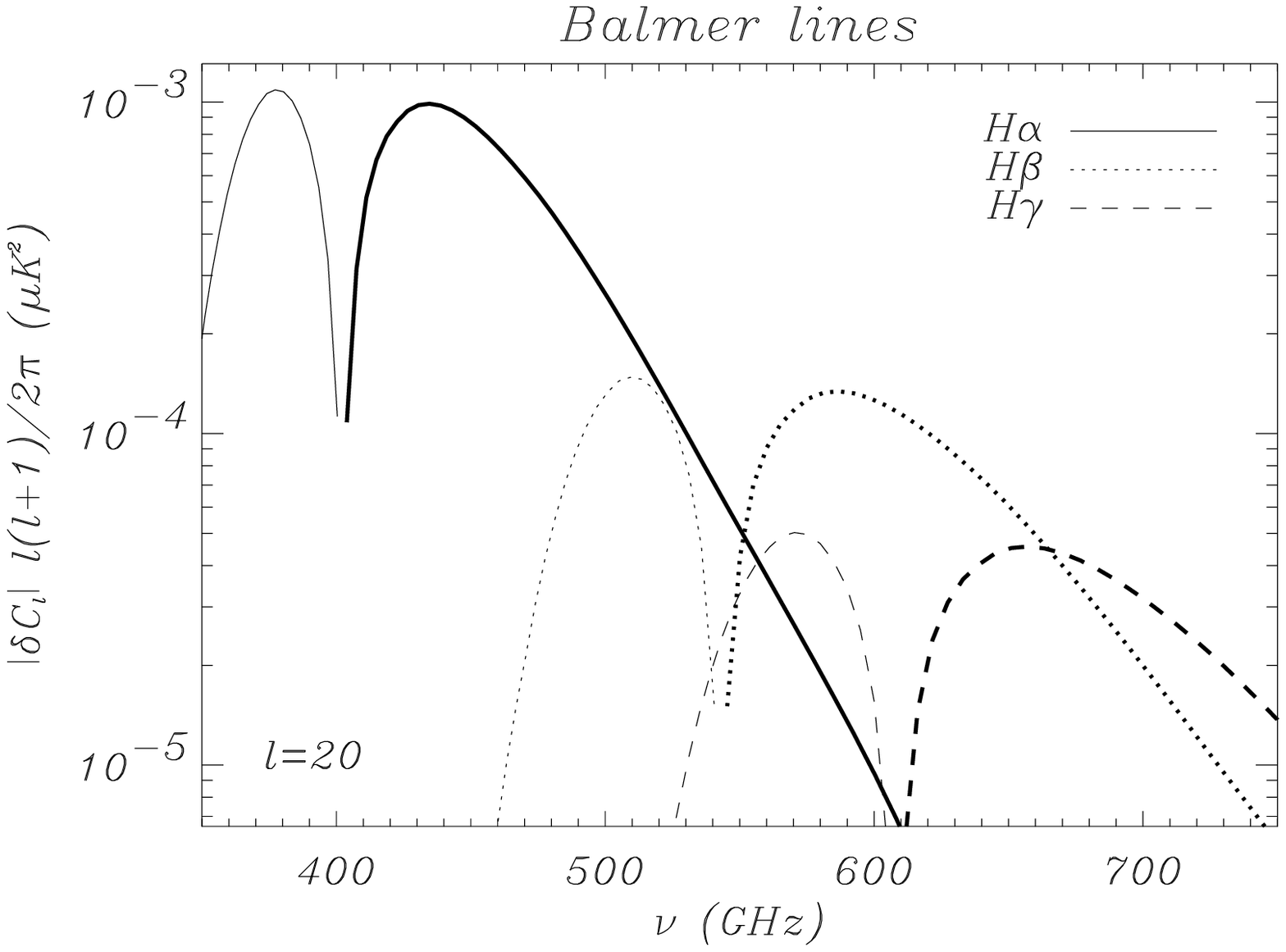}
\includegraphics[width=0.8\columnwidth]{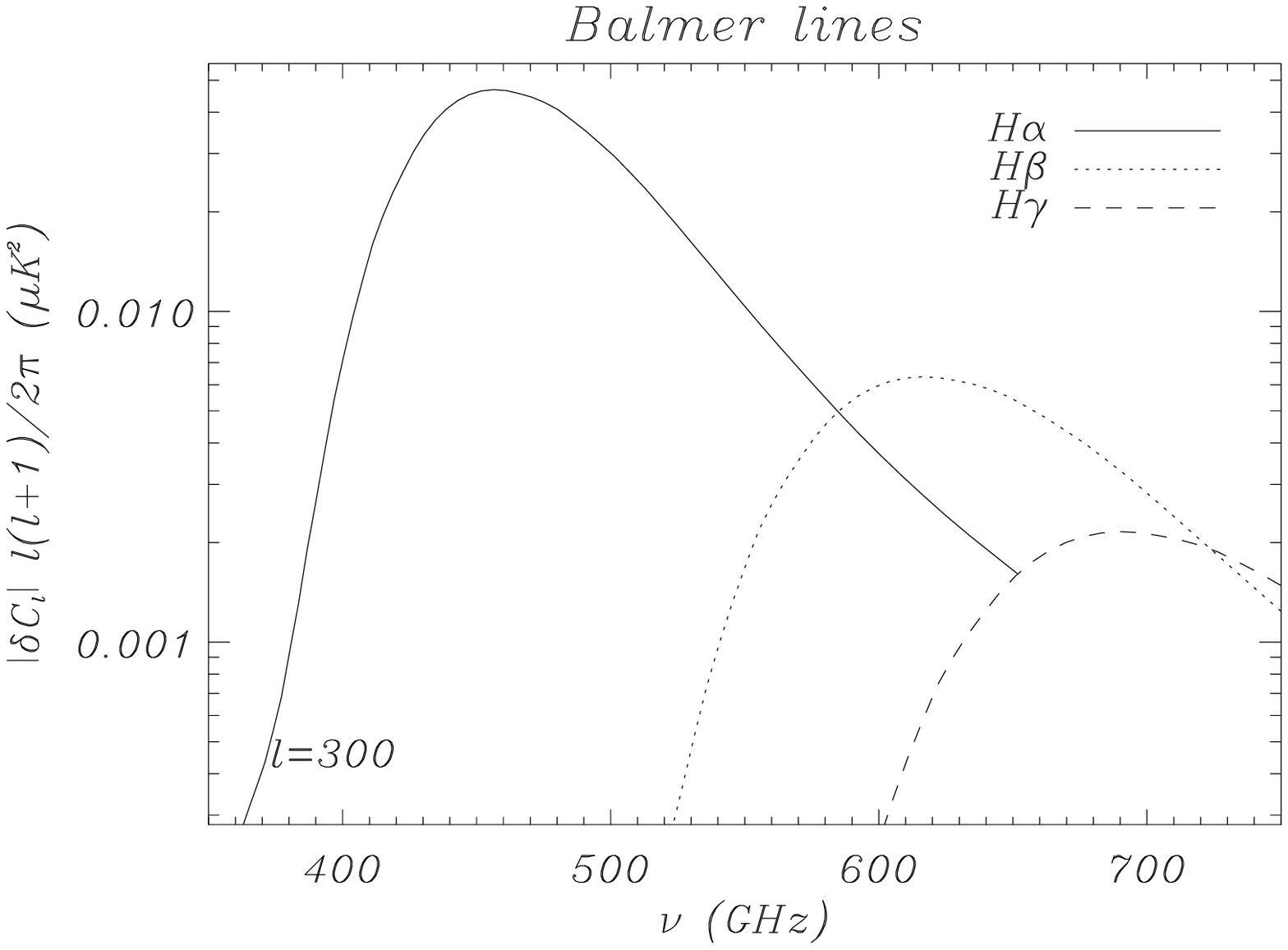}
\includegraphics[width=0.8\columnwidth]{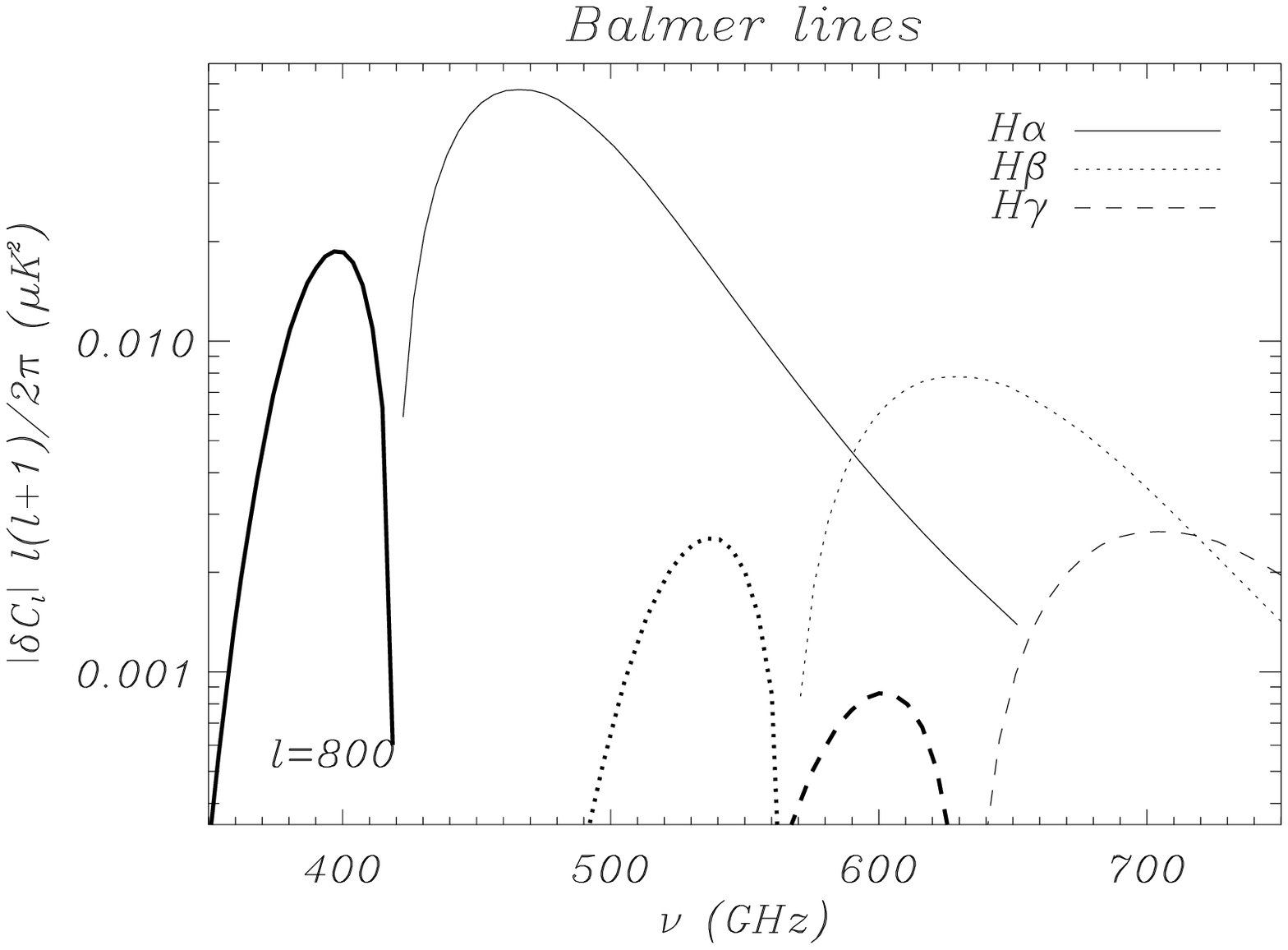}
\caption{Angular power spectrum arising from the Balmer recombination lines
of hydrogen during cosmological recombination, as a function
of the redshifted (observed) frequency. We plot the values
at $\ell=20$ (top panel), at $\ell=300$ (middle panel), and at 
$\ell=800$ (bottom panel). Thick lines corresponds to region where the
$\delta C_\ell$ are positive, whereas thin lines correspond to
negative values. }
\label{fig:balmer}
\end{figure}

\begin{figure}
\includegraphics[width=\columnwidth]{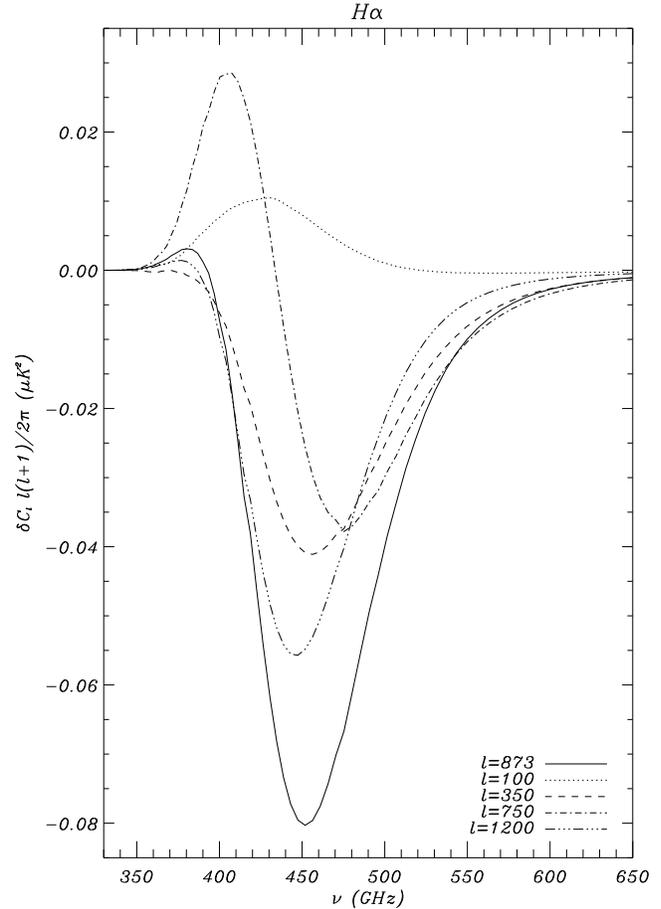}
\caption{Same as Figure~\ref{fig:balmer}, but only for the H$\alpha$
line which gives the maximum contribution. The vertical axis is now
presented in a linear scale so we can explicitly see the change of
sign. We plot values for five multipole scales, $\ell=100, 350, 750,
1200$, and also $\ell=873$, which is the multipole at which we have 
the maximum signal at $\nu=452$~GHz. [See movie in the electronic version].  }
\label{fig:features}
\end{figure}

We now consider the frequency dependence of the effect. 
Figure~\ref{fig:balmer} shows the relative 
change in the power spectrum due to the presence of the hydrogen 
recombination lines of Balmer series, as a function of the 
present day redshifted frequency. 
For each Balmer line, we show three different panels corresponding to
three different angular scales ($\ell = 20, 300$ and $800$). 
As pointed above, the signal has a very peculiar frequency dependence,
and for example at angular scales of $\ell=20$ and $\ell=800$, we can
see a characteristic change of sign of the effect, related to the
positions of the cancellations of the generation and blurring terms. 
For the most intense line, H$\alpha$, we show five more angular scales 
in Fig.~\ref{fig:features}, where we explicitly use a linear scale 
in the vertical axis to show the change of sign. The case of
$\ell=873$, which is giving the maximum signal (in absolute value) is 
shown as a solid line in that figure. 
The frequency dependence of the effect 
is unique and completely different from that
of foregrounds (e.g. dust emission), and thus it might 
be used to separate these signals from other contaminants, as we will
see below. 

\begin{figure*}
\includegraphics[width=\columnwidth]{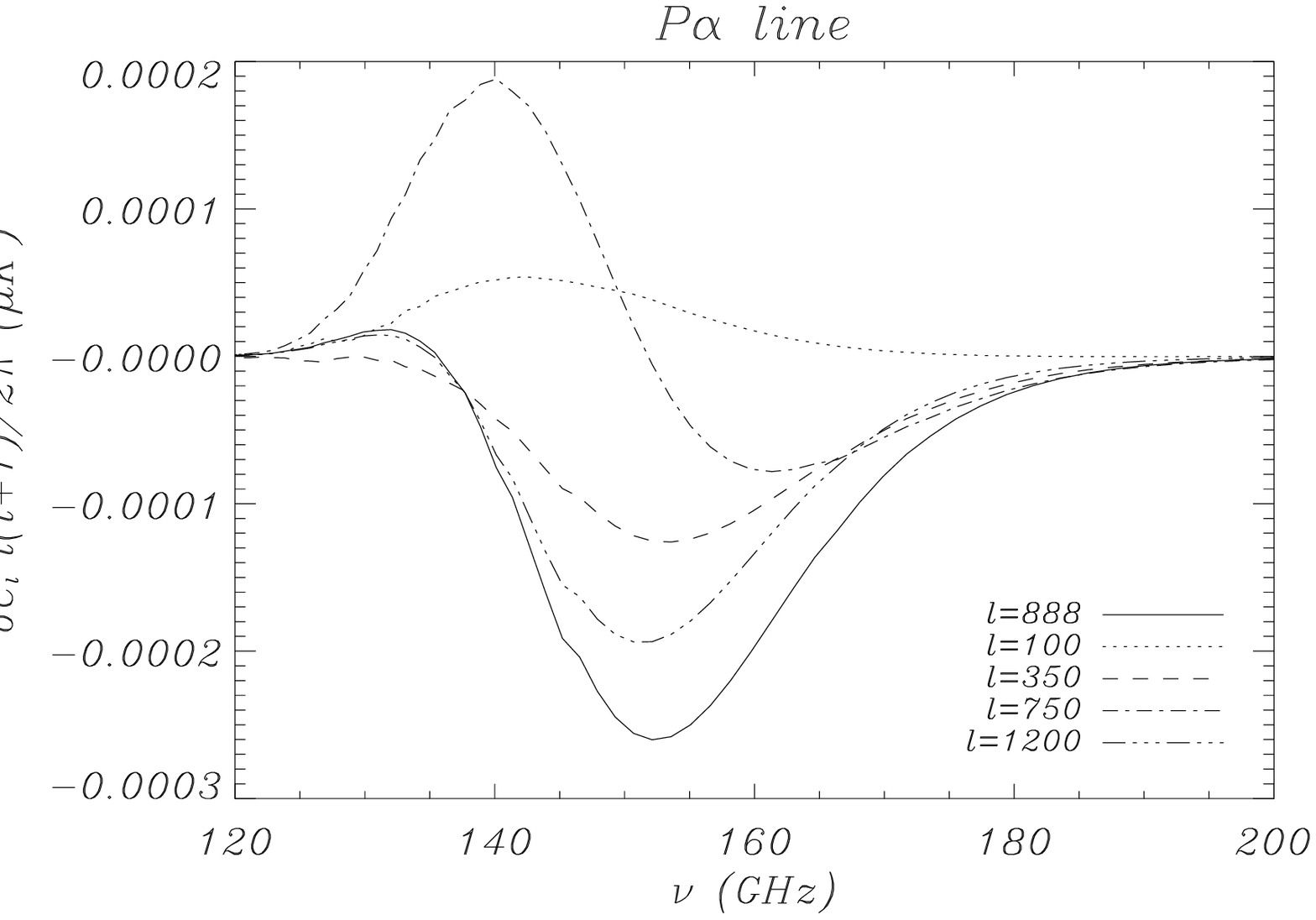}%
\includegraphics[width=\columnwidth]{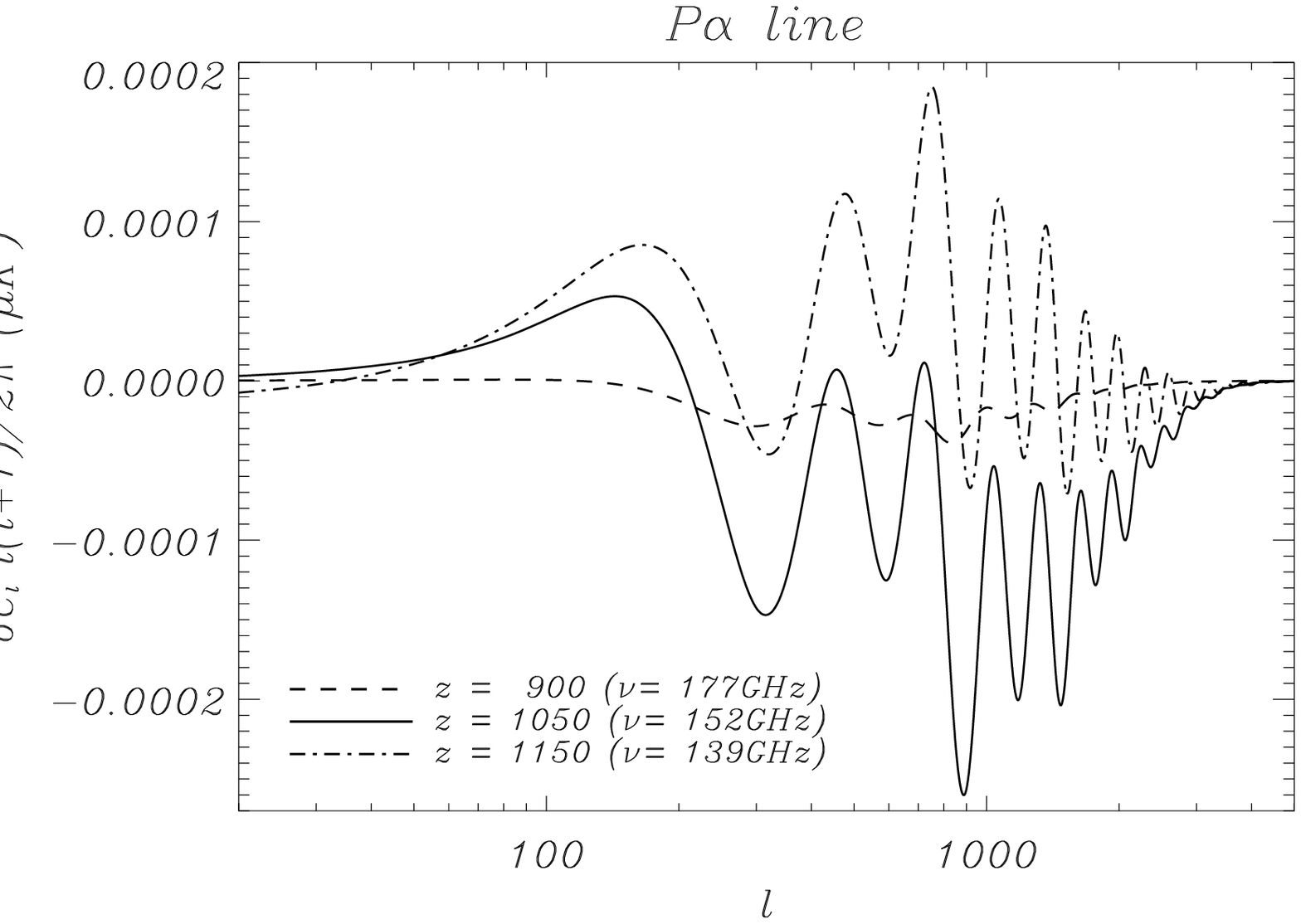}
\includegraphics[width=\columnwidth]{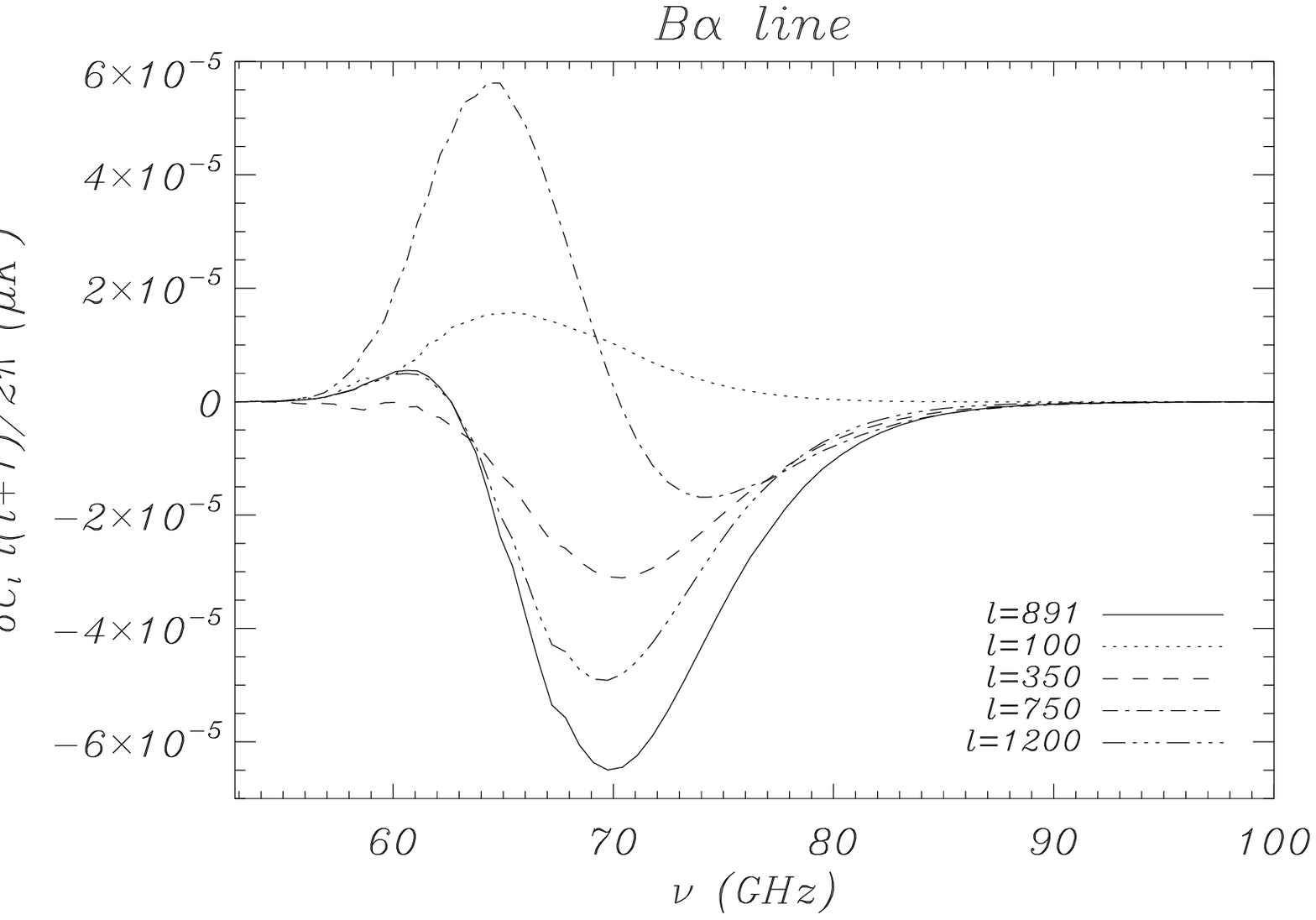}%
\includegraphics[width=\columnwidth]{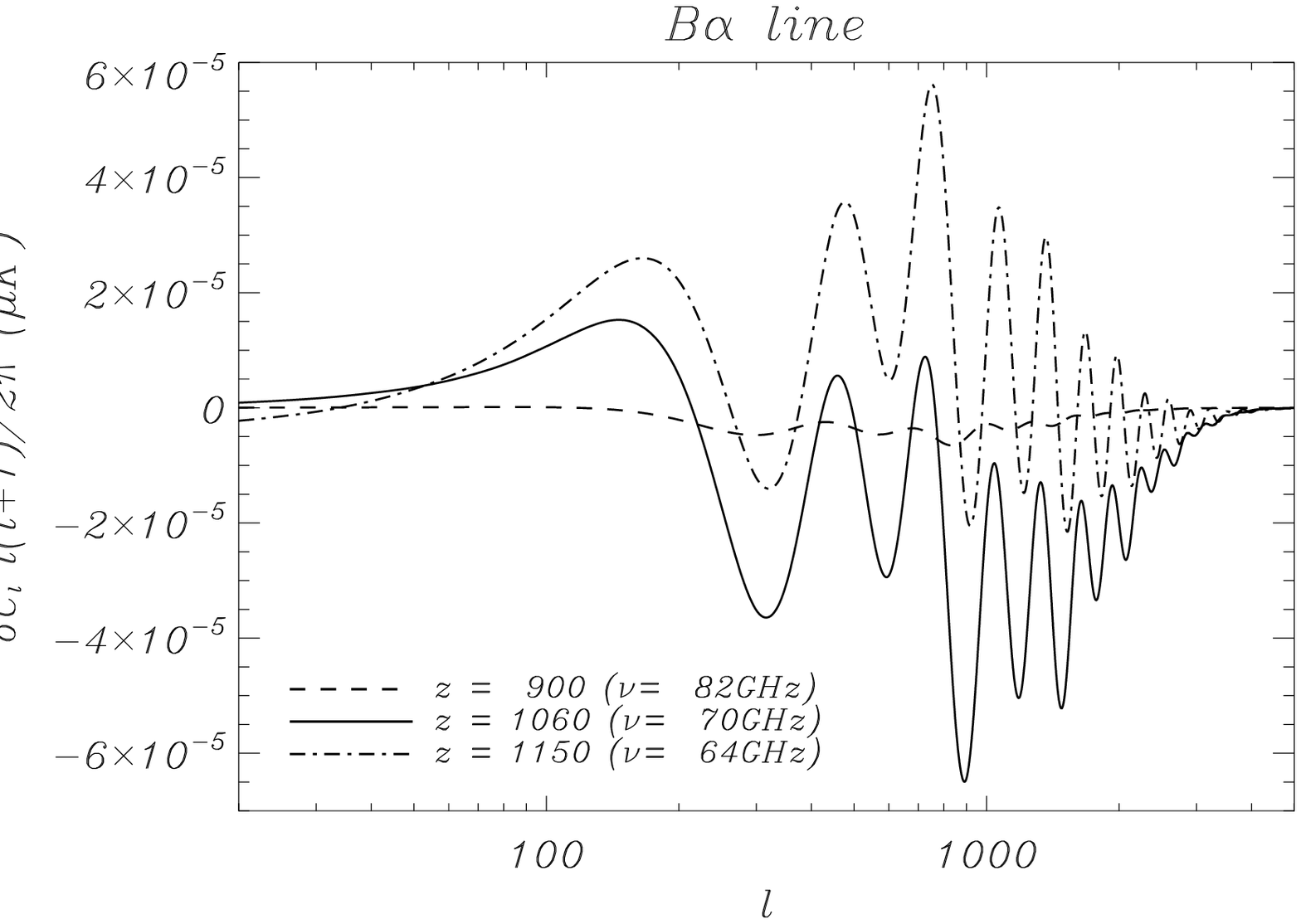}
\caption{Angular power spectrum arising from the P$\alpha$ (top 
row) and B$\alpha$ (bottom row) hydrogen lines during recombination, as
a function of the redshifted frequency (left column) and the angular
multipole (right column). The solid lines in all panels refer to the cases 
in which the signal is maximum, according to Table~\ref{tab:maximos}. }
\label{fig:paschen_and_brackett}
\end{figure*}

The figures for the Paschen and Brackett lines 
are similar to those for Balmer lines. 
For illustration, in Figure~\ref{fig:paschen_and_brackett} 
we present the angular and
frequency dependence for the B$\alpha$ and the P$\alpha$ lines.
The amplitude of the signal for all the transitions considered in this
paper is presented in Table~\ref{tab:maximos}. 

Summarizing this subsection, we conclude that 
the best strategy to observe these 
features and perform this
tomography would be to use bandwidths ($\Delta \nu/\nu$) 
of the order of 10\%, although bandwidths of 1\% would trace this
signal much better. 
These values are narrower than the width of the 
visibility function in the line, but still wide enough to keep enough photons.
It is clear that the major contribution to the effect during the epoch
of recombination is coming from high $\ell$ multipoles, and that
the widths of the features is of the order of $\Delta \ell \sim 140$,  
so this opens the possibility to use not only satellites but 
balloons or ground based experiments to look for this signal.

\section{Contamination from other effects}

In this section we discuss how the presence of foregrounds and the effect
associated with Rayleigh scattering might affect the study of the lines. 
We shall adopt the {\rm middle-of-the-road} model of
\citet[hereafter T00]{2000ApJ...530..133T}, 
and we consider the contribution of 
five foreground sources, namely synchrotron radiation, free-free
emission, dust emission, thermal SZ effect associated
with filaments and superclusters of galaxies, and Rayleigh scattering. 
Point sources are not considered here, so we assume that resolved
sources can be extracted from the maps, and that the contribution of
unresolved sources can be lowered down to roughly the noise level of
the observing instrument. 

For the first three components, synchrotron, free-free and dust, 
the angular dependence of the power spectra was approximated by a
power law. The frequency dependence for free-free and synchrotron was
taken to be a power law, whereas a modified black-body spectrum was
used for the dust component. 
For the SZ effect, the frequency dependence is well-known. 
Details of the modeling of the 
angular dependence can be found in T00. This 
neglects the contribution from the SZ effect generated by
resolved clusters of galaxies, which we assume can be removed
from the maps.
For the Rayleigh scattering effect modeling, we follow BHMS04. 
As discussed above, the scattering cross-section of this process has a
very strong dependence on the frequency, 
\[
\sigma_R(\nu) = \sigma_T \nu^4 \Bigg( \sum_{k\ge2} \frac{f_{1k}}%
{(\nu_{1k}^2 - \nu^2)} \Bigg)^2 \propto 
\Bigg( \frac{\nu}{\nu_{12}} \Bigg)^4
\]
where the last proportionality holds for $\nu \ll \nu_{12}$. Thus, the net
effect on the power spectrum will also show this dependence 
($\delta C_\ell^{R} \propto \tau_{R} \propto \nu^4$). Unfortunately,
this frequency dependence mimics the spectrum of dust emission from
local dust and bright extragalactic star-forming galaxies, so the
separation of these effects could be difficult. 

\begin{figure*}
\includegraphics[width=2\columnwidth,height=10cm]{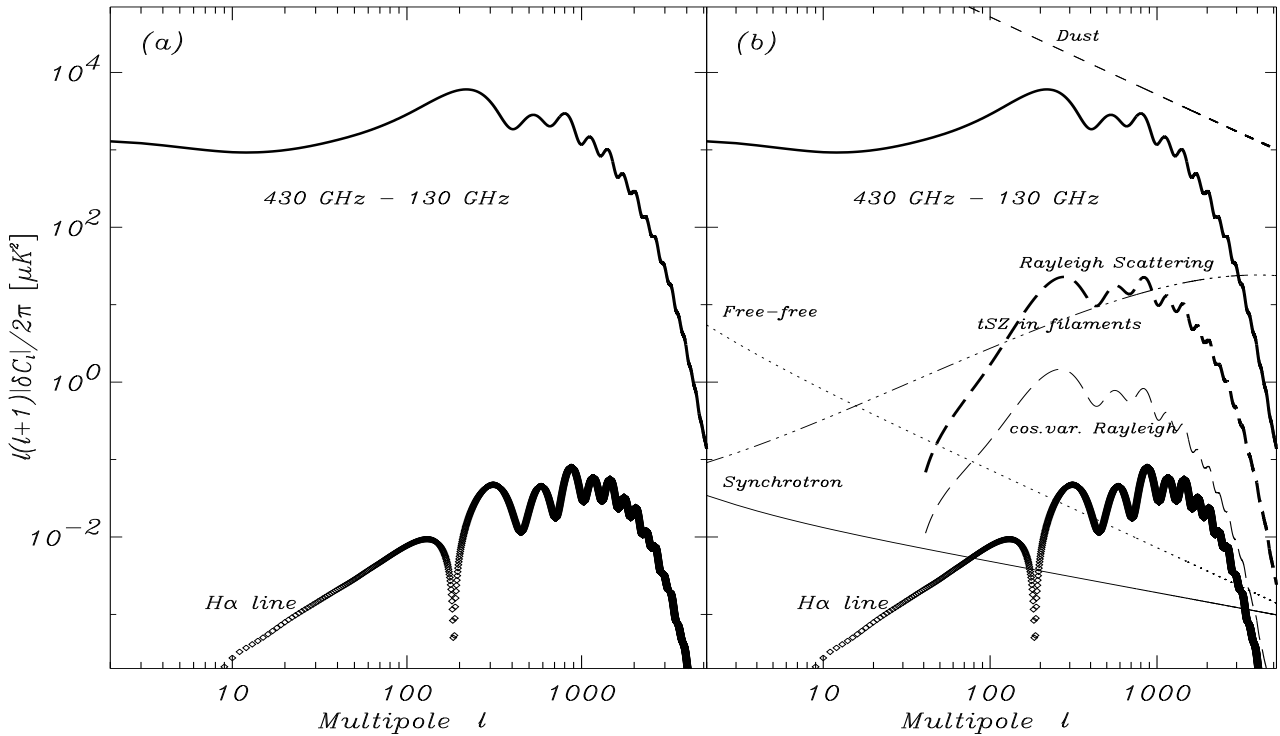}
\caption{(a) Maximum contribution of the coherent scattering in the 
H$\alpha$ line to the angular CMB power spectrum, as seen by an 
hypothetical experiment measuring 
at 430~GHz and 130~GHz. The upper thick solid line gives the reference
power spectrum. 
(b) Contribution of foregrounds at these frequencies. All thin lines
refer to the foreground model as quoted in T00: free-free emission is
shown as dotted line; synchrotron emission as solid line; the thermal
SZ effect produced in filaments as dot-dashed line and the thermal
dust emission is shown by dashed lines. Rayleigh scattering 
contribution is shown as a thick long-dashed line, whereas the cosmic
variance contribution associated to this effect is shown as a thin
long-dashed line.  }
\label{fig:foregrounds}
\end{figure*}

Fig.~\ref{fig:foregrounds} shows the expected level, prior to any
removal, of the discussed foregrounds when looking for the maximum of
the H$\alpha$ line. We considered here the case of an ideal experiment 
measuring at 430~GHz and 130~GHz, and with an instrumental bandwidth
of $\Delta \nu/\nu=10^{-2}$, which is comparable to the electron
width of the line under discussion. 
It is clear that  the main diffuse contamination will come
from dust emission, but there are other two signals which will be
larger than the H$\alpha$ contribution, namely 
the thermal SZ emission from filaments and the Rayleigh scattering.
For the first one, the signal has a completely different
frequency and angular behavior from the one we are considering, so
these peculiarities could be used to separate both components. 

\begin{figure}
\includegraphics[width=\columnwidth]{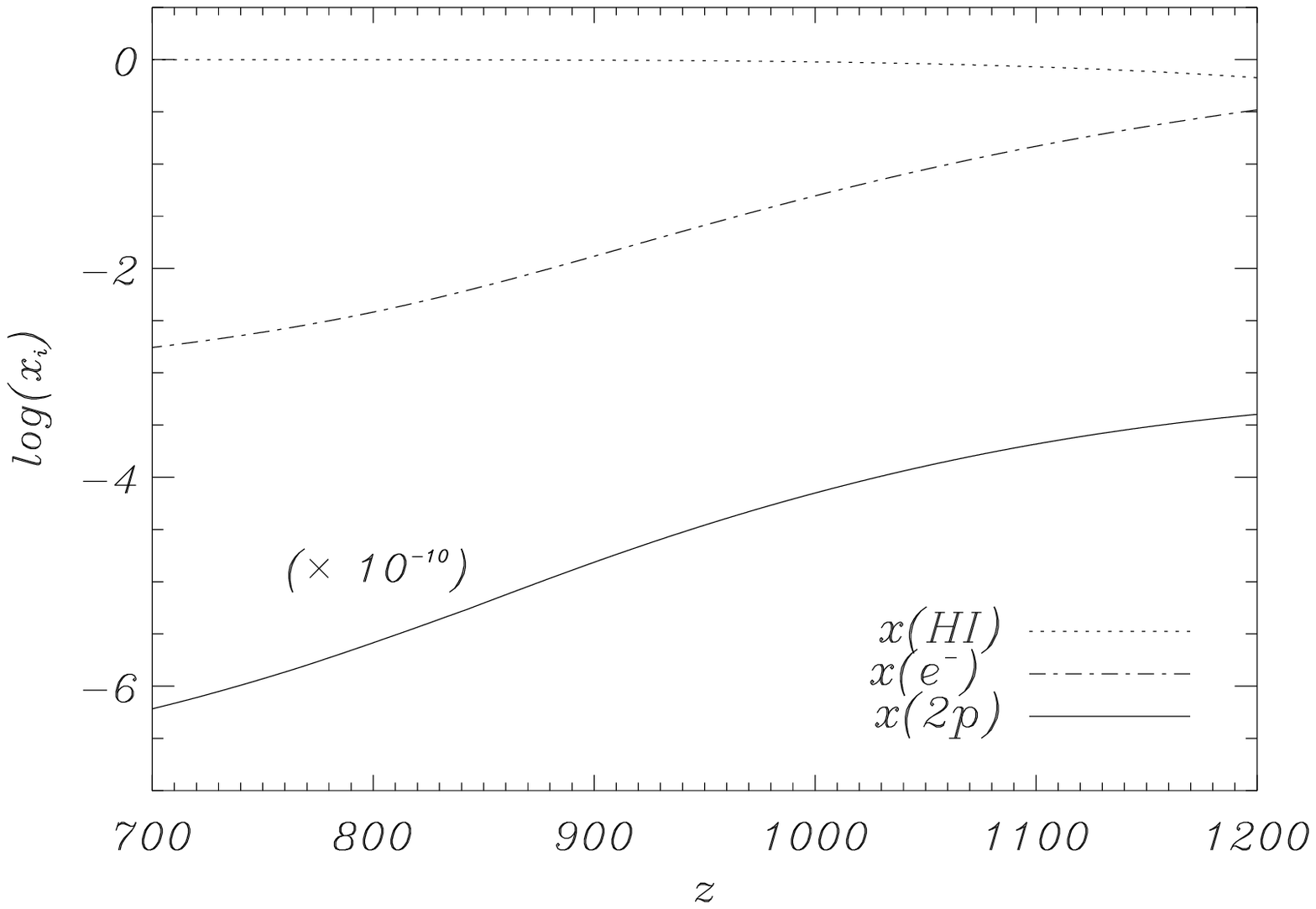}
\caption{Multilevel hydrogen recombination computed in this paper. It
is presented the evolution of the relative fraction of 
electrons ($x_e$), neutral hydrogen atoms ($x_{HI}=n_{HI}/n_{H}$) and
the population of the 2p level ($x_{2p}=n_{2p}/n_H$), as a function of
the redshift $z$ in the vicinity of the epoch of recombination (near
the peak of the visibility function). The $x_{2p}$ values were multiplied by
$10^{10}$ to display all lines in the same scale. 
The cosmological model adopted
here is the same as in Fig.~\ref{fig:x_e}.}
\label{fig:jalberto}
\end{figure}

For the case of Rayleigh scattering, the frequency behavior is also 
completely different from the case of resonant scattering. 
The angular dependence has some similarities in the high multipole
range (damping tail) if we probe a line in the low redshift tail 
of recombination, because here both effects are proportional to 
$-2\tau C_\ell$. However, as we probe higher redshifts, the 
behaviors become different also in this domain.
In addition to this, 
there is another difference between the Rayleigh scattering and
the coherent scattering in lines, which is related to the
different evolution of the optical depth of each process during recombination,
and which is superimposed on the previous effect. 
To illustrate it, 
we show in Fig.~\ref{fig:jalberto} the redshift dependence of the
population of electrons, neutral hydrogen atoms and 
the population of hydrogen atoms in the level 2p, all
computed using our code. These three variables ($n_e$, $n_{HI}$ and
$n_{2p}$) are proportional to 
the differential optical depth $\dot{\tau}$ for the Thomson, Rayleigh
and coherent scatterings, respectively. 
From this, it is clear that 
the redshift interval $\delta z/z$ in which the optical depth for 
Rayleigh scattering 
changes significantly is much larger than the corresponding one of
the H$\alpha$ line scattering. Thus, instruments with broad bandwidths 
will dilute the signal from the lines, but add up the signal from
Rayleigh scattering. This also could be used to separate them, because
using a broad-band instrument we can isolate the Rayleigh scattering
component, and then subtract it. 

One interesting physical remark is that 
the evolution of the population of the 2p
level in Fig.~\ref{fig:jalberto} is practically proportional to 
the electron density in this redshift range. 
This is what we expected, as we can immediately see from 
Eq.~\ref{saha} ($n_{2p} \propto n_e^2 \exp \{ \chi_2/(k_B T_e) \}$) and 
from the evolution of the electron density 
($x_e \propto (1/z) \exp \{ -\chi_2 / (k_B T_e) \}$, see ZKS68).

\section{Discussion and conclusions}

We have studied here the imprint of cosmological hydrogen recombination
lines on the power spectrum of the CMB. To this end, we have developed
a code that follows the evolution of the population of the levels of
the hydrogen atom, separately treating those levels with different
angular momentum. 
From here, we have obtained the optical depths associated with coherent
scattering in these lines during the epoch of recombination, and these
numbers were used to compute the effect on the angular power spectrum
of the CMB. 

These changes are small ($0.1\mu K$-$0.3\mu K$), but could be
separated from other effects due to their peculiar frequency and
angular dependence. 
Unfortunately, the line giving the maximum signal (the H$\alpha$
line) is placed in a frequency domain where the contribution from dust
emission, the tSZ effect produced in filaments and Rayleigh scattering
are important, so it will be necessary either to look for
regions with low contamination (for the case of dust), 
or to use component separation techniques to reach this signal. 
The important point here is that the signal under discussion has a 
very characteristic behavior, both in frequency and in angular scale,
which is completely different from any of the above
contaminants. 

One of the most important properties of this signal is that each
observing frequency is associated with one redshift, so observations of
these signals at different frequencies might give information 
about the amplitude of the fluctuations and corresponding peculiar 
velocities of matter at different redshifts during the epoch of recombination.

In this paper, the best-fit cosmological model to
the WMAP 1st-year data \citep{2003ApJS..148....1B} was 
adopted to make predictions about
the shape and the intensity of the signal we expect to measure. 
With the announced sensitivities of future experiments, these signals
should be detected, and thus we could try to infer some information
on the values of the cosmological parameters.
The inferred constraints from the detection of these lines 
will be clearly independent of those obtained using the
model-fitting to the power spectrum of the CMB.
This additional information could be used
to break some degeneracies in the parameters in two ways. 
First, as pointed out above, observations in narrow spectral bands
of these signals can give us information about specific redshift
slices during the process of recombination in the Universe. 
The shape and positions of the peaks and zeros of the signal 
presented in Figs.~\ref{fig:features2} and \ref{fig:deltaCl_to_cl} 
reveal the details of the baryon transfer function at each 
considered epoch, so they could be used as an additional test of the 
cosmological model. 
For a given experiment with a given bandwidth, it is straight-forward to derive
the prediction for $\delta C_\ell$, so model fitting could be used here, in the
same way as it is used today to extract parameters from the power spectrum. 
Secondly, the determination of the amplitude of this effect 
at a given redshift could also be used to set an independent constraint on 
$\Omega_b h^2$.

Using this signal, future narrow-band spectral observations 
might permit one to study all the details of the transfer 
function describing the evolution and viscous damping 
of perturbations in the epoch of recombination.

\acknowledgements{%
We acknowledge use of the CMBFast software package \citep{cmbfast}.
JARM and CHM acknowledge the financial support provided through the European
Community's Human Potential Programme under contract HPRN-CT-2002-00124, 
CMBNET.}\\

\label{lastpage}

\end{document}